\documentclass[fleqn,usenatbib]{mnras}

\usepackage{newtxtext,newtxmath}
\usepackage{ulem}

\usepackage[T1]{fontenc}
\usepackage{ae,aecompl}
\usepackage{color}

\usepackage{graphicx}	
\usepackage{amsmath}	
\usepackage{amssymb}	

\defcitealias{Parikh2018}{P18}
\defcitealias{Johansson2012}{J12}
\defcitealias{ThomasD2011b}{TMJ}

\title[Chemical abundance gradients]{SDSS-IV MaNGA: local and global chemical abundance patterns in early-type galaxies}

\author[T. Parikh et al.]{
Taniya Parikh$^{1}$\thanks{E-mail: taniya.parikh@port.ac.uk},
Daniel Thomas$^{1}$,
Claudia Maraston$^{1}$,
Kyle B. Westfall$^{2}$,
Jianhui Lian$^{1}$,
\newauthor{Amelia Fraser-McKelvie$^{3}$,
Brett H. Andrews$^{4}$,
Niv Drory$^{5}$,
Sofia Meneses-Goytia$^{1,6}$
}
\\
$^{1}$Institute of Cosmology and Gravitation, University of Portsmouth, 1-8 Burnaby Road, Portsmouth PO1 3FX, UK\\
$^{2}$UCO/Lick Observatory, University of California, Santa Cruz, 1156 High St. Santa Cruz, CA 95064, USA\\
$^{3}$School of Physics and Astronomy, University of Nottingham, University Park, Nottingham, NG7 2RD, UK\\
$^{4}$PITT PACC, Department of Physics and Astronomy, University of Pittsburgh, Pittsburgh, PA 15260, USA\\
$^{5}$McDonald Observatory, The University of Texas at Austin, 1 University Station, Austin, TX 78712, USA\\
$^{6}$Department of Physics, University of Surrey, Guildford GU2 7XH, UK
}

\date{Accepted XXX. Received YYY; in original form ZZZ}

\pubyear{2018}

\begin{document}
\label{firstpage}
\pagerange{\pageref{firstpage}--\pageref{lastpage}}
\maketitle

\begin{abstract}
Chemical enrichment signatures strongly constrain galaxy formation and evolution, and a detailed understanding of abundance patterns provides clues regarding the nucleosynthetic production pathways of elements. Using the SDSS-IV MaNGA IFU survey, we study radial gradients of chemical element abundances in detail. We use stacked spectra out to $1\;R_e$ of 366 early-type galaxies with masses $9.9 - 10.8\;\log M/M_{\odot}$ to probe the abundances of the elements C, N, Na, Mg, Ca, and Ti, relative to the abundance of Fe, by fitting stellar population models to a combination of Lick absorption indices. We find that C, Mg, and Ti trace each other both as a function of galaxy radius and galaxy mass. These similar C and Mg abundances within and across galaxies set a lower limit for star-formation timescales. Conversely, N and Ca are generally offset to lower abundances. The under-abundance of Ca compared to Mg implies delayed enrichment of Ca through Type Ia supernovae, whereas the correlated behaviour of Ti and the lighter $\alpha$ elements, C and Mg, suggest contributions to Ti from Type II supernovae. We obtain shallow radial gradients in [Mg/Fe], [C/Fe], and [Ti/Fe], meaning that these inferences are independent of radius. However, we measure strong negative radial gradients for [N/Fe] and [Na/Fe], of up to $-0.25\pm0.05$ and $-0.29\pm0.02$~dex/$R_e$ respectively. These gradients become shallower with decreasing galaxy mass. We find that N and Na abundances increase more steeply with velocity dispersion within galaxies than globally, while the other elements show the same relation locally and globally. This implies that the high Na and N abundances found in massive early type galaxies are generated by internal processes within galaxies. These are strongly correlated with the total metallicity, suggesting metallicity-dependent Na enrichment, and secondary N production in massive early-type galaxies.
\end{abstract}




\section{Introduction}
The chemical compositions of stellar atmospheres are tracers of element abundances of the parent gas clouds forming stars throughout the formation history of a galaxy, providing a powerful tool for extracting information on galaxy formation and evolution as well as the chemical enrichment history of the Universe. As the most prominent example, the ratio between $\alpha$-elements, like O, Mg, Si, and Fe-peak elements constrains star formation timescales in galaxies because of the delayed enrichment of iron from Type Ia supernovae \citep{Matteucci1986, Thomas1999}. As [$\alpha$/Fe] increases with increasing galaxy mass for early-type galaxies \citep[e.g.][]{Trager2000,Thomas2005, Thomas2010, Bernardi2006, Clemens2006}, massive galaxies are expected to have formed on shorter timescales than low-mass ones \citep{Matteucci1994, Thomas2005, Thomas2010}. Alternative explanations for this trend could include a top-heavy IMF or selective galactic winds \citep{Thomas1999}, although evidence for this has not been found.

Individual element abundance ratios, in addition to [$\alpha$/Fe], can further disentangle the formation of different stellar populations. As an example, the balance between the elements carbon and magnesium sets a lower limit on the star formation time-scale, as star formation must continue over long enough periods to allow for the contribution of C from both massive and intermediate-mass stars \citep[][J12]{Johansson2012}. The element nitrogen, mostly produced in intermediate-mass stars, provides further constraints on formation timescales, but also primordial gas inflow (because of primary vs secondary N production) \citepalias{Johansson2012}. However, we are far from a comprehensive understanding of these processes; for instance, chemical enrichment models struggle to explain C and N abundances in early-type galaxies \citep{Pipino2009}.

Other elements such as Na, Ca, Ti are produced in a variety of processes at different timescales during the chemical evolution of a galaxy including AGB star winds, Type Ia and Type II supernovae \citep{Conroy2014} and can therefore be used to further constrain the physics of galaxy formation and evolution. In some cases the nucleosynthesis origins also need to be understood better.

Most work in the literature so far has focused on central, integrated properties \citep[e.g.][]{Thomas2005, Schiavon2007, Graves2007, Thomas2010, Johansson2012, Conroy2014} with only a handful of spatially resolved studies in the most recent literature \citep{Greene2015, vanDokkum2016, Alton2018}. The largest spatially resolved sample for which individual element abundances are determined is the study of 100 early type galaxies by \citet{Greene2015}, for a limited set of elements (C, N, Mg, Ca) and focusing on massive galaxies.

Recent advances in spatially resolved spectroscopy have paved the way to carry out such studies for a more representative set of the local galaxy population. It is important to understand whether global trends found among galaxy populations also hold locally within galaxies. Such a comparison provides fundamental clues for galaxy evolution as well as the origin of chemical element production.



In this paper, we aim to study detailed chemical abundance patterns within galaxies, making use of spatially resolved spectroscopy from the survey Mapping Nearby Galaxies at Apache Point Observatory \citep[MaNGA][]{Bundy2015}. High resolution spectroscopy for a large sample of nearby galaxies, with the necessary spatial and wavelength coverage, provides the opportunity to contribute to previous efforts of extragalactic archaeology. In \citet[][P18]{Parikh2018}, we described our data sample, the stacking procedure to produce high quality, high S/N spectra, and determined radial gradients in the low-mass end IMF slope of 366 early type galaxies. Now, we make use of absorption features in the optical region of the same data, to derive chemical abundance ratios for all accessible elements, which are C, N, Na, Mg, Ca, and Ti. We probe these using nine optical indices and stellar populations models from \citet{ThomasD2011b}, present the ages, metallicities, and abundances, and focus particularly on how these trends vary radially within galaxies and globally with galaxy mass.

The paper is structured as follows. In Section~2 we give a brief description of the data and refer the reader to the relevant sections in \citetalias{Parikh2018}. We describe the additional indices and parameters studied in this work. Our radial gradients are presented in Section~3, as well as analysis of local versus global relations. We compare our results to literature in Section~4 and discuss the possible origins of chemical elements and the driver of abundance variations, followed by our conclusions and scope for further work in Section~5.

\section{Data and analysis tools}
MaNGA, part of the Sloan Digital Sky Survey IV \citep{Blanton2017}, aims to obtain spatially resolved spectroscopy for 10,000 nearby galaxies at a spectral resolution of $R\sim 2000$ in the wavelength range $3,600-10,300\;$\AA, upon completion in 2020. It conducts 17 simultaneous observations of galaxies from the output of independent fibre-bundles \citep{Drory2015}, which are fed into the BOSS spectrographs \citep{Smee2013} on the Sloan $2.5\;$m telescope \citep{Gunn2006}. MaNGA targets are chosen from the NASA Sloan Atlas catalogue \citep[NSA,][]{Blanton2005} such that there is a uniform distribution in log stellar mass \citep{Wake2017}. We make use of the $r$-band isophotal ellipticity, position angle, galaxy mass \citep[based on a Chabrier IMF,][]{Chabrier2003}, and half-light radius from this catalogue, always using Elliptical Petrosian quantities.

Optical fibre bundles of different sizes are chosen to ensure all galaxies are covered out to at least $1.5 R_\mathrm{e}$ for the `Primary' and `Color-enhanced' samples, and to $2.5 R_\mathrm{e}$ for the `Secondary' sample \citep{Wake2017}. The Color-enhanced sample supplements colour space that is otherwise under-represented relative to the overall galaxy population. The spatial resolution is $1 - 2\;$kpc at the median redshift of the survey ($z\sim 0.03$), and the $r$-band S/N is $4-8\;$\AA$^{-1}$, for each 2\arcsec\ fibre, at the outskirts of MaNGA galaxies. For more detail on the survey we refer the reader to \citet{Law2015} for MaNGA's observing strategy, to \citet{Yan2016a} for the spectrophotometry calibration, to \citet{Wake2017} for the survey design, and to \citet{Yan2016b} for the initial performance.

We make use of the LOGCUBE files from the Data Reduction Pipeline \citep[DRP,][]{Law2016}, where the wavelength vector has been logarithmically binned. Briefly, the DRP processes all exposures taken for each galaxy into datacubes, involving spectral extraction, flux calibration, and subtracting sky lines and continuum from the raw observed spectra. Information is recorded as a function of wavelength at each on-sky sample, giving spatial pixels (spaxels). MaNGA's Data Analysis Pipeline (DAP, Westfall et al., in prep) fits the stellar continuum and nebular emission lines of each spectrum, providing the kinematics of both components, as well as emission-line fluxes and equivalent widths. We use the stellar velocities and velocity dispersions from the DAP for each spectrum in our analysis.

\begin{figure*}
  \includegraphics[width=.95\linewidth]{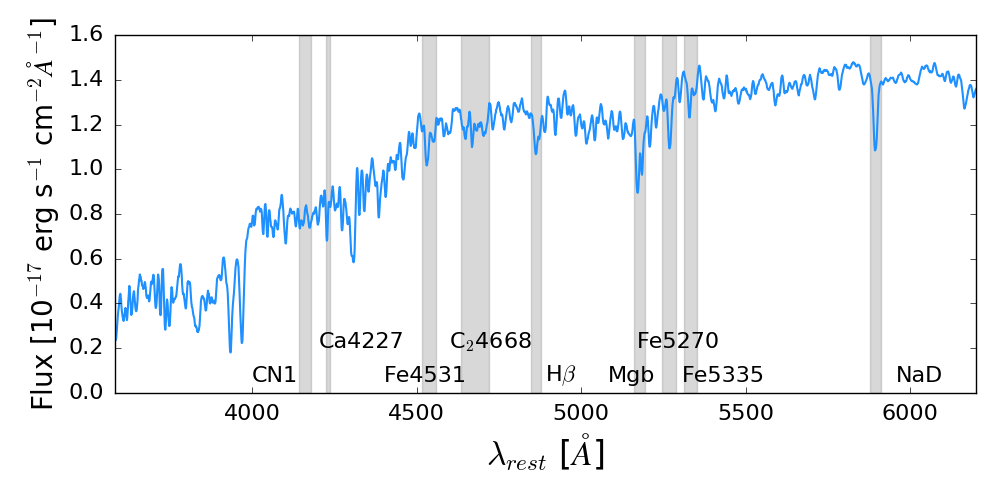}
\caption{An example stacked spectrum is shown from our intermediate mass bin. The indices used in this work are labelled and highlighted in grey, for the radius 0.4 - 0.5 R$_e$.}
\label{fig:example}
\end{figure*}
\subsection{Stacked spectra sample}
Section 2.1 in \citetalias{Parikh2018} describes in detail our galaxy sample and stacking procedure. We provide only a brief description below. Starting from data taken during the first two years of survey operations, equivalent to SDSS's fourteenth data release \citep[DR14]{Abolfathi2017} containing $2812$ datacubes, we select early-type galaxies using Galaxy Zoo morphologies \citep{Lintott2011, Willett2013}, visually inspecting them when necessary \citep{Goddard2017}. We apply a S/N cut of ${\rm S}/{\rm N}=7$\; pixel$^{-1}$ on individual MaNGA spectra so that we only include spectra with reliable stellar velocities. We stack spectra in radial bins for individual galaxies, and then stack galaxies together. Stacking increases S/N, mitigates problems due to sky residuals and allows studies out to larger radii.

The bins in radius are from 0-0.1 $R_e$ to 0.9-1.0 $R_e$ in steps of 0.1 $R_e$ giving ten bins, where the effective radius $R_e$ is defined as the half-light radius of the galaxy. In the final stacks, we only include galaxies which have at least one spectrum in each radial bin after the S/N cut, and we cut the tails of low mass and high mass galaxies to provide a uniform distribution in mass. The resulting three mass bins, containing 122 galaxies each, are centred on $\log M/M_{\odot}$ of 10, 10.4, and 10.6, and a central velocity dispersion $\sigma$ of 130, 170, and $200\;$km/s, respectively.

An example stacked spectrum is shown in Fig.~\ref{fig:example} from our intermediate mass bin. We choose an intermediate radial bin and label the indices used in this work, highlighted as grey shaded regions.

\subsection{Spectral analysis}
\label{sec:abs_ind}
Extracting stellar population parameters from spectra can be done via full spectrum fitting or absorption index fitting \citep[for a review, see][]{Conroy2013}. Literature regarding chemical abundances has widely focused on analysing indices (e.g. \citealp{Sanchez-Blazquez2003, Graves2007, Thomas2010, Price2011}; \citetalias{Johansson2012}) and we adopt this method since our high S/N stacked spectra allow precise measurements of absorption features.

We determine stellar kinematics and subtract emission lines from an initial full spectrum fit using the DAP's pPXF wrapper (\citealt{Cappellari2004}; Westfall et al., in prep). The DAP $\sigma$ values for each spectrum are propagated for the stacked spectra and used as a guess for the full spectrum fit. These account for the difference between the instrumental resolution of models and data, as well as the astrophysical velocity dispersion of the galaxies. As discussed in \citetalias{Parikh2018} the $\sigma$ profiles derived from the fit are unphysically flat, likely because of the larger uncertainties in radial velocities at larger radii, which effectively become another dispersion convolved with the spectrum. We therefore only use these derived $\sigma$ values for resolution matching of models and data. Following our approach in \citetalias{Parikh2018}, median values of the DAP $\sigma$ measurements from the individual spectra, are instead adopted whenever physically meaningful velocity dispersions are discussed.

Indices consist of continuum bandpasses on either side of a feature bandpass and are calculated using equations 1-3 from \citet{Worthey1994}. The definitions \citep[from][]{Trager1998} are shown in Table~\ref{tab:indices} for all the indices we make use of in this work. The method to measure the indices, to apply velocity dispersion corrections, and to calculate Monte Carlo-based errors is detailed in \citetalias[][section 2.3.2]{Parikh2018}. Briefly, we use the MILES-based stellar population models from \citet{Maraston2011} to measure the index strengths two times, once at the original resolution of the model and once at the resolution of the data. The latter is produced by convolution of the model spectra with both the instrumental resolution and the astrophysical velocity dispersion. The ratio between the two measurements provides us with our correction factor. We note here that the magnitude-unit index CN1 used in this work is calculated slightly differently from the other indices and its velocity dispersion-correction factor is additive rather than multiplicative.

\begin{table}
	\centering
	\caption{Index definitions for the absorption features used in this work \citet{Trager1998}. All wavelengths are in vacuum; indices are measured in units of \AA, except for CN1, which is measured in magnitude units.}
	\label{tab:indices}
	\begin{tabular}{lccc}
		\hline
		Index & Blue Continuum & Feature & Red continuum\\
		\hline
		CN1 & 4081.3 -  4118.8 & 4143.3 -  4178.3 &  4245.3 - 4285.3\\
		Ca4227 & 4212.2 -  4220.9 & 4223.4 -  4235.9 &  4242.2 - 4252.2\\
		Fe4531 & 4505.5 -  4515.5 & 4515.5 -  4560.5 &  4561.8 - 4580.5\\
		C$_2$4668 & 4612.8 -  4631.5 & 4635.3 -  4721.6 &  4744.1 - 4757.8\\
		H$\beta$ & 4829.2 - 4849.2 & 4849.2 - 4878.0 & 4878.0 - 4893.0\\
		Mg$b$ & 5144.1 - 5162.8 & 5161.6 - 5194.1 & 5192.8 - 5207.8\\
		Fe5270 & 5234.6 - 5249.6 & 5247.1 - 5287.1 & 5287.1 - 5319.6\\
		Fe5335 & 5306.1 - 5317.4 & 5313.6 - 5353.6 & 5354.9 - 5364.9\\
		NaD & 5862.2 - 5877.3 & 5878.5 - 5911.0 & 5923.8 - 5949.8\\
		\hline
	\end{tabular}
\end{table}
We derive abundance ratios [X/Fe] for X = C, N, Na, Mg, Ca, and Ti using carefully chosen indices which are sensitive to each of these element abundances. We use the same prescription as in \citetalias{Johansson2012}, where the individual elements are varied on $\alpha$ enhanced models. Hence,
\begin{equation}
[{\rm X}/{\rm Fe}] = [{\rm X}/\alpha] + [\alpha/{\rm Fe}].
\end{equation}

The choice of indices is based on \citetalias[][Fig.~1]{Johansson2012}. We adopt a simplistic approach whereby the number of indices used is kept minimal, similar to \citetalias{Parikh2018}. We use C$_2$4668 to determine C; CN1 for N; NaD for Na; Mgb for Mg; Ca4227 for Ca; and finally Fe4531 to measure Ti. This set of indices is more limited than \citetalias{Johansson2012} because we find that no further information is gained by using other indices (e.g. CN2, Mg$_1$, Mg$_2$) and we want to avoid using IMF-sensitive indices (e.g. TiO1, TiO2).

Although each of these indices is most sensitive to the element we are using it to measure, some also respond to the presence of other elements. CN1 is also sensitive to [C/Fe] and Ca4227 is also sensitive to both [C/Fe] and [N/Fe]; hence we model these indices after independently determining the other parameters they are sensitive to \citep[similar to][]{Greene2015}. E.g. we use C$_2$4668 to determine and fix C, before using CN1 to measure N, and similarly Ca4227 is modelled after both C and N have been determined. As presented in \citet[][figure 1]{Johansson2012}, Fe4531 does not respond to O in the TMJ models. We note, however, that other models suggest that Fe4531 is sensitive to O as well as Ti (G. Worthey, private communication). This caveat should be kept in mind when using  Fe4531 as Ti abundance indicator.

In the final step, we model all indices with all the derived parameters, to show that the data is consistent with the ages, total metallicities, and individual element abundance variations.

\begin{figure*}
  \flushleft
  \includegraphics[width=.32\linewidth]{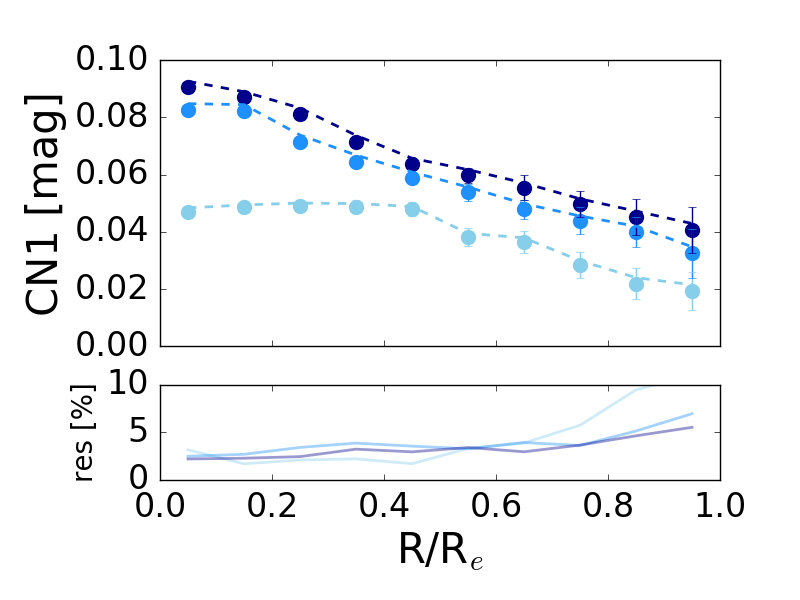}
  \includegraphics[width=.32\linewidth]{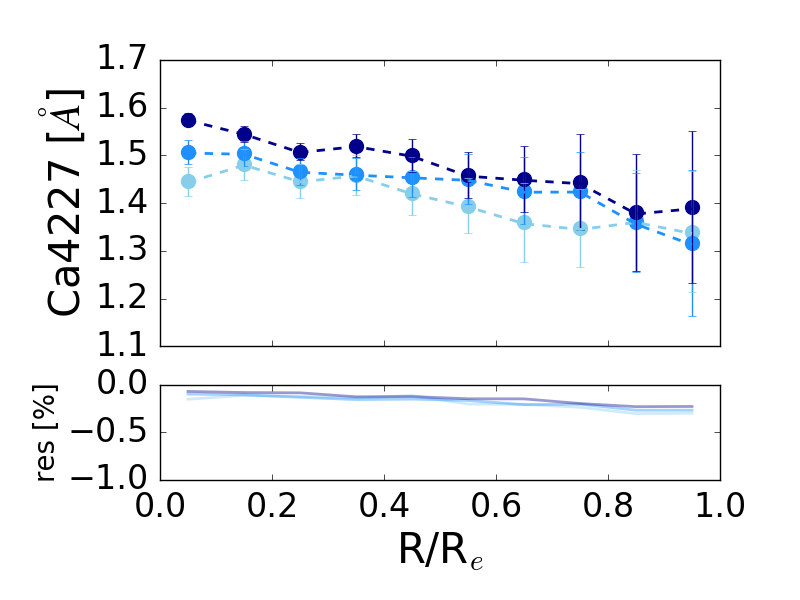}
  \includegraphics[width=.32\linewidth]{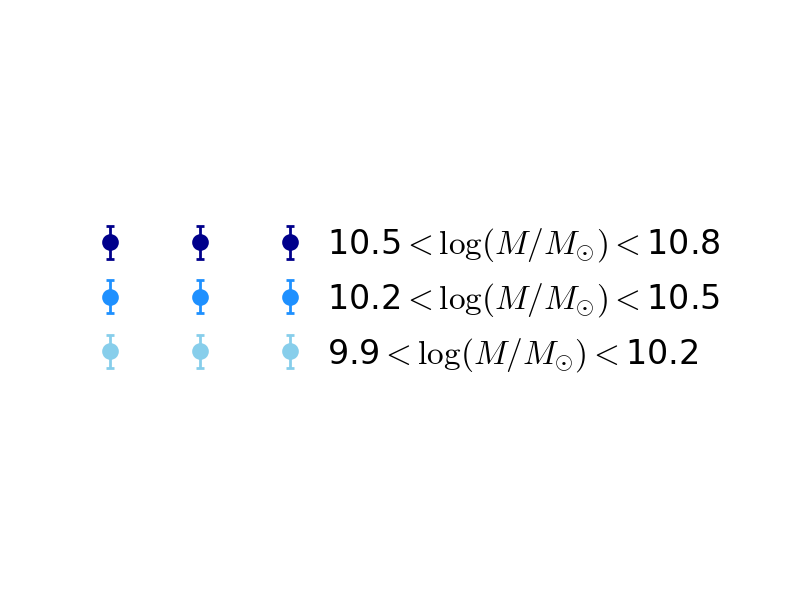}
  \includegraphics[width=.32\linewidth]{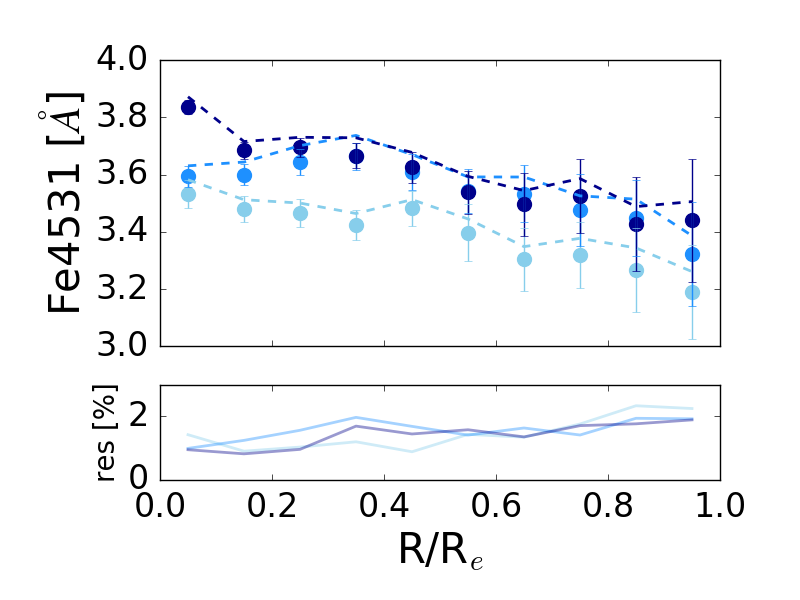}
  \includegraphics[width=.32\linewidth]{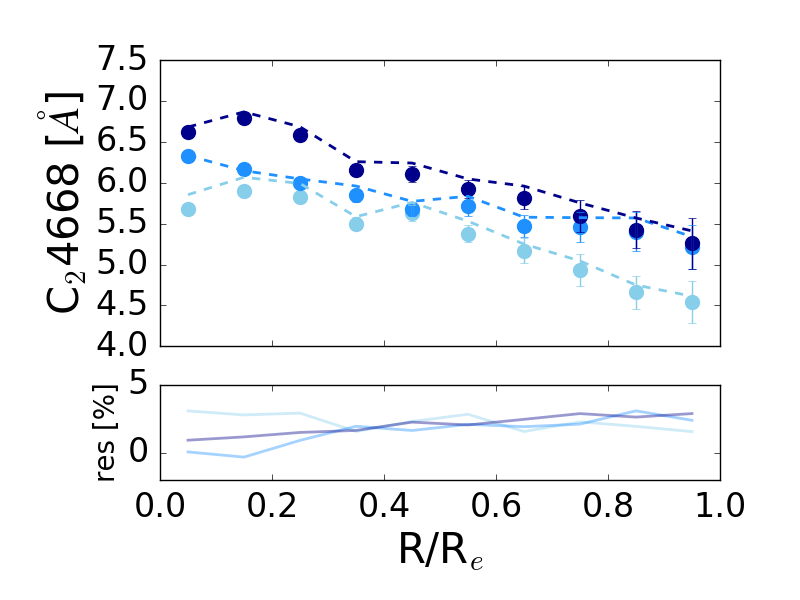}
  \includegraphics[width=.32\linewidth]{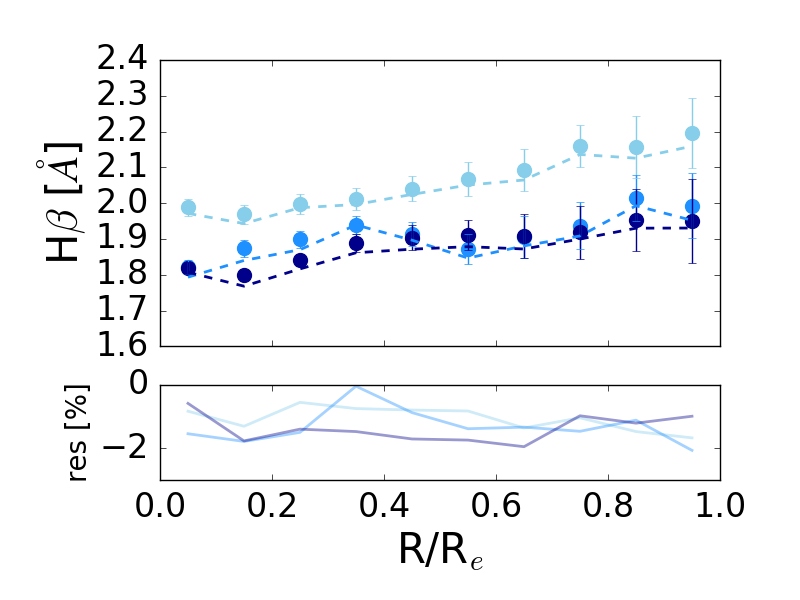}
  \includegraphics[width=.32\linewidth]{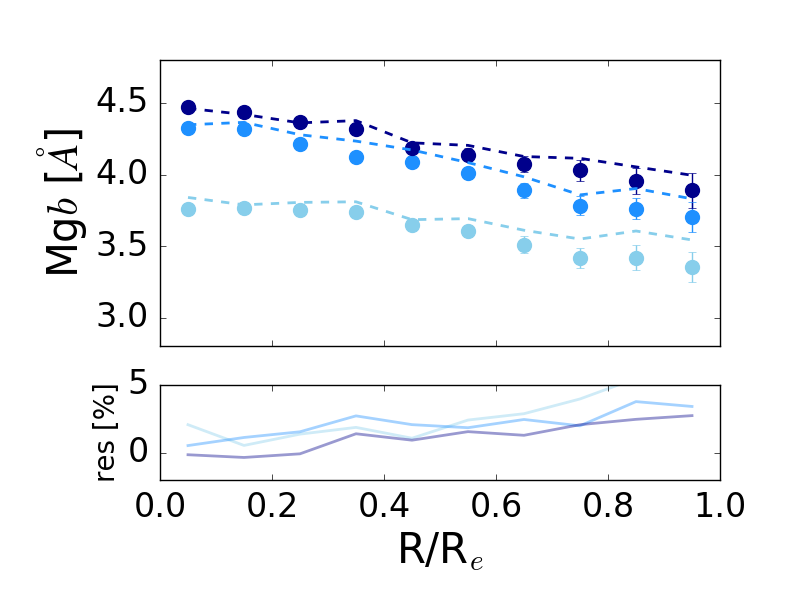}
  \includegraphics[width=.32\linewidth]{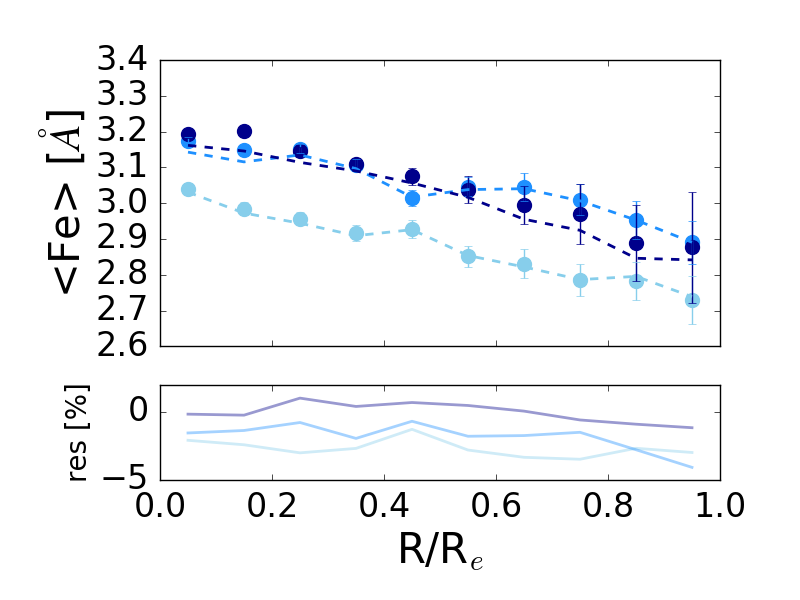}
  \includegraphics[width=.32\linewidth]{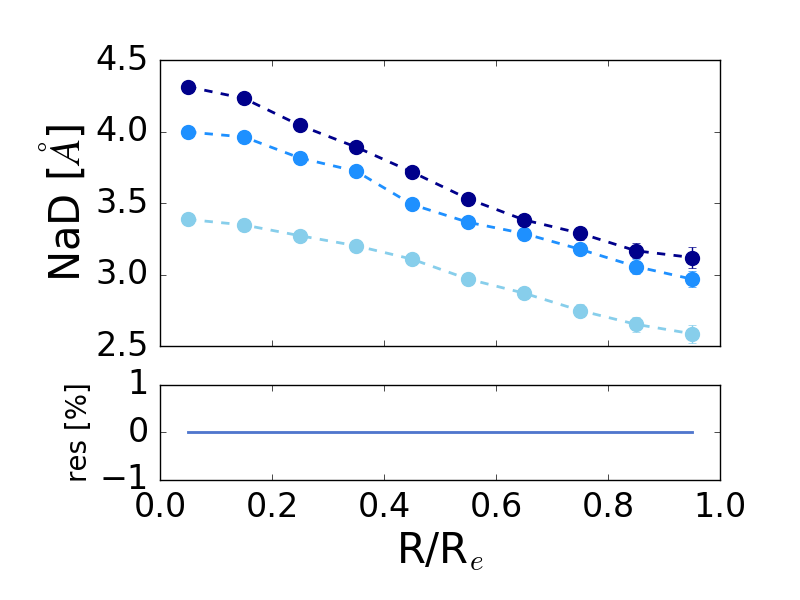}
\caption{CN1, Ca4227, Fe4531, C$_2$4668, H$\beta$, Mg$b$, <Fe>, and NaD for each mass bin as a function of radius. Line-strengths (circles) are measured on the stacked spectra with 1-$\sigma$ errors calculated using a Monte Carlo-based analysis. The dashed lines are \citetalias{ThomasD2011b} model fits obtained through chi-squared minimisation (see text). The residuals from the fits are shown at the bottom of each panel. Measurements and models are both shown at MILES resolution.}
\label{fig:indices}
\end{figure*}

\subsection{Models}
\label{sec:models}
We use the \citet[][TMJ]{ThomasD2011b} models of the optical Lick absorption indices, which are an update and extension of the earlier models by \citet{Thomas2003,Thomas2004}, and based on the \citet{Maraston1998,Maraston2005} evolutionary synthesis code. These models use empirical calibrations by \citet{Johansson2010} based on the MILES stellar library \citep{Sanchez-Blazquez2006}, and element response functions from \citet{Korn2005}. The models are carefully calibrated with galactic globular clusters and reproduce the observations well for the spectral featured used in this study \citep{Thomas2011a}.

The models are available for different ages, metallicities, variable element abundance ratios, and a Salpeter \citep{Salpeter1955} IMF. Element variations are calculated at constant total metallicity, hence the \citetalias{ThomasD2011b} models enhance $\alpha$ and other light elements (C, N, O, Na, Mg, Si, Ca and Ti), and suppress the Fe-peak elements according to Equations 1-3 in \citet{Thomas2003}. We use the models provided at MILES resolution of $2.5\;$\AA\ \citep{Beifiori2011,Falcon-Barroso2011,Prugniel2011} in the present work, which matches the MaNGA resolution to a good approximation over the relevant wavelength range.

\section{Results}
\label{sec:results}
We now present the index strength measurements, which are modelled in steps using the \citetalias{ThomasD2011b} models described in Section~\ref{sec:models} to derive stellar population parameters. We make use of H$\beta$, Mg$b$, Fe5270, and Fe5335 as age, metallicity and [$\alpha$/Fe] indicators, similar to \citetalias{Parikh2018}. These parameters are derived simultaneously by fitting stellar population models to data using a chi-squared minimisation code. To this end, the models are interpolated in a 3-dimensional parameter space. We then fix these parameters and individually vary element abundances in order to reproduce the relevant indices. The IMF is fixed because its effect on the optical indices used in this work is negligible.

In \citetalias{Parikh2018} we use [$\alpha$/Fe] and [Mg/Fe] interchangeably; since [$\alpha$/Fe] is derived using the four optical indices without taking individual element ratios into account, it is equivalent to Mg abundance \citepalias{Johansson2012}. Hence in this work we focus on [Mg/Fe] rather than [$\alpha$/Fe].

\subsection{Index gradients and stellar population model fits}
The absorption indices are shown as circles in Fig.~\ref{fig:indices} as a function of galaxy radius. The two iron indices, Fe5270 and Fe5335, are averaged to give <Fe>. It can be seen that indices are stronger in more massive galaxies with the exception of H$\beta$ which displays an anti-correlation. H$\beta$ increases with radius; all other indices display negative radial gradients, with gradients in CN1, NaD, and Mg$b$ becoming steeper with increasing mass.

The errors on the indices are calculated using a Monte Carlo-based approach. In all cases the error increases with radius. Certain spectral regions, including the Mg$b$ and NaD features, have very low error on the flux in the stacked spectra, hence the errors on the index measurements are smaller than the symbols on this plot.

Stellar population parameters and element abundances are derived by modelling the indices in steps, as described before. The model predictions are shown as dashed lines in Fig.~\ref{fig:indices} and include the combination of age, total metallicity, and individual element abundances derived at each radial bin. The corresponding model parameters are presented in the following subsection. The indices and models are both at the MILES resolution. The residuals are shown at the bottom of each panel. The models fit the data well; the residuals are of the order of a a few percent or less. The index measurements and the derived parameters with errors are provided in Tables~\ref{tab:ind_measurements} \& \ref{tab:parameters} in the Appendix for all mass and radial bins.

\begin{figure*}
  \includegraphics[width=.32\linewidth]{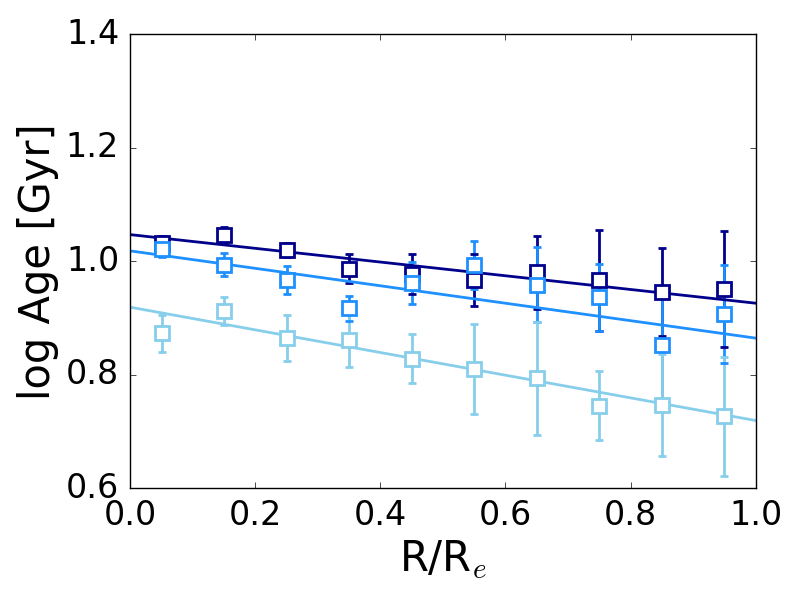}
  \includegraphics[width=.32\linewidth]{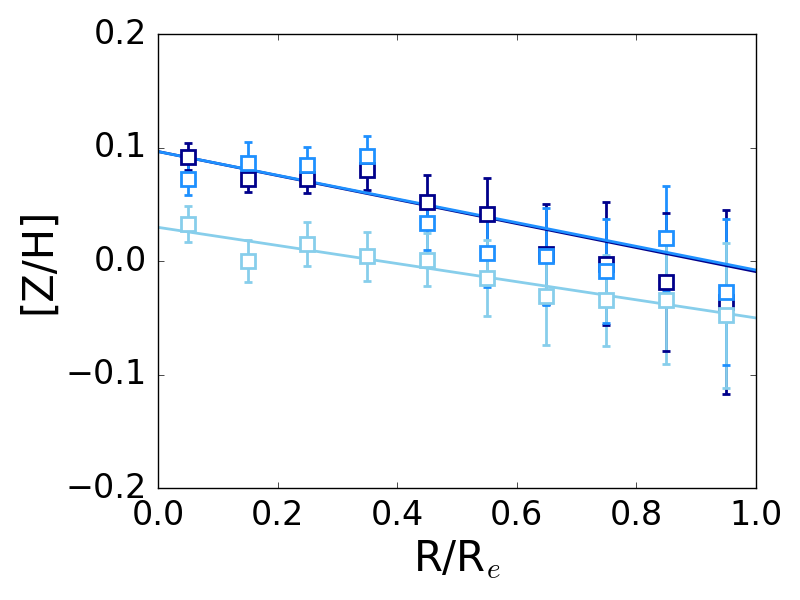}
  \includegraphics[width=.32\linewidth]{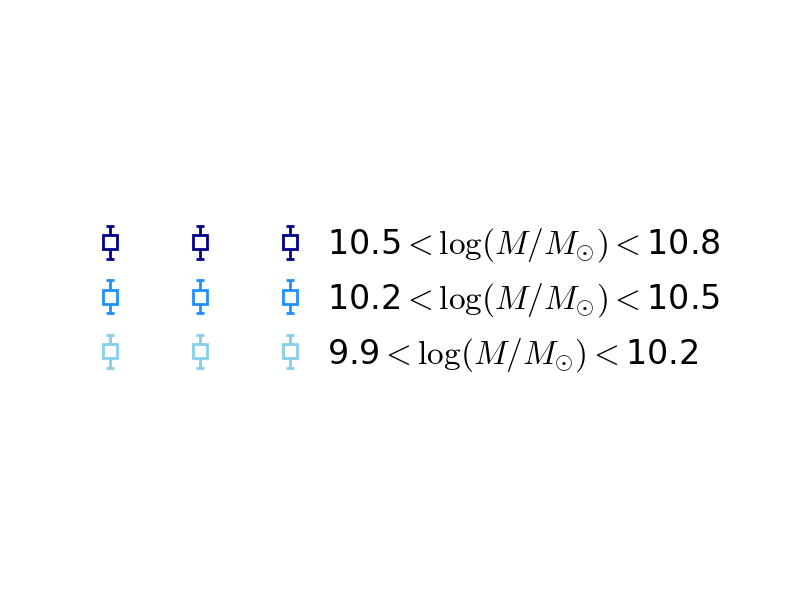}
  \includegraphics[width=.32\linewidth]{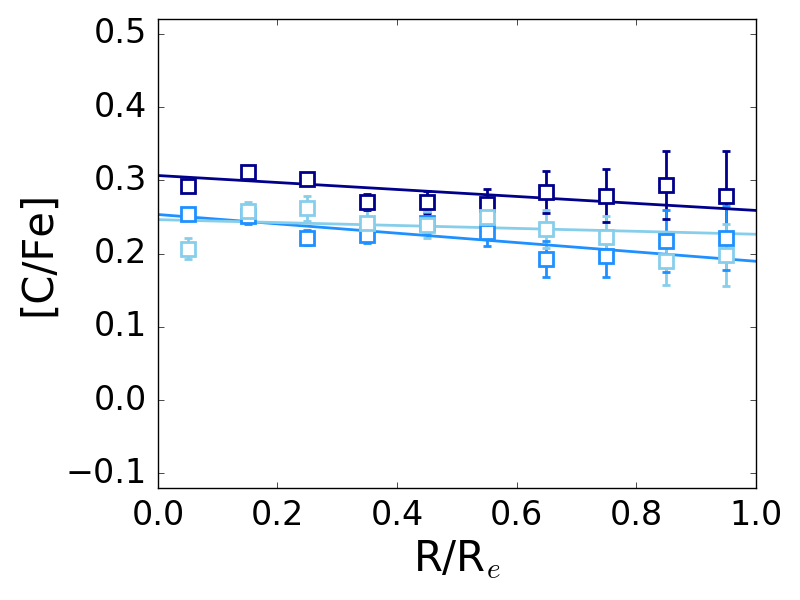}
  \includegraphics[width=.32\linewidth]{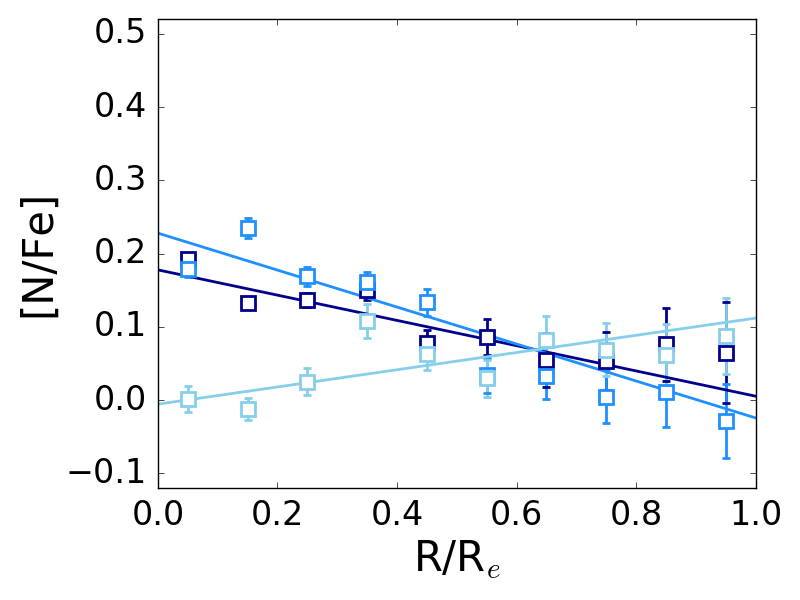}
  \includegraphics[width=.32\linewidth]{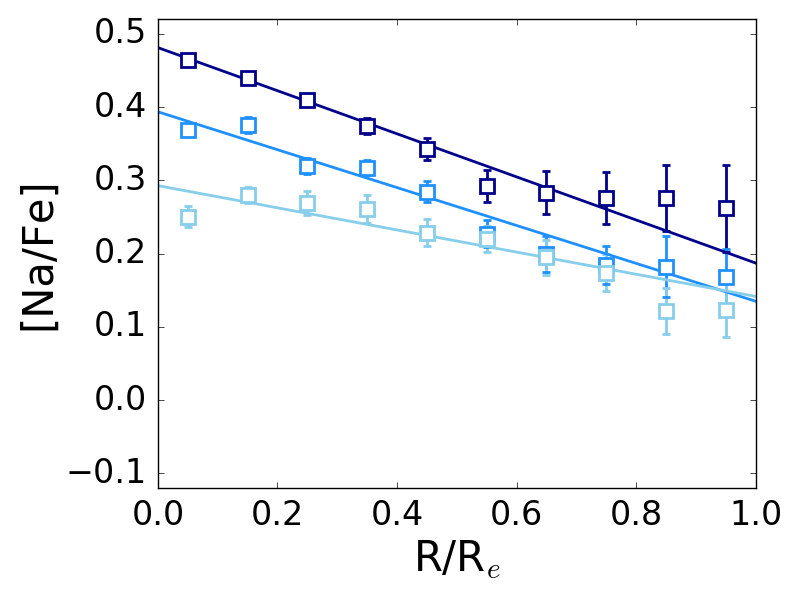}
  \includegraphics[width=.32\linewidth]{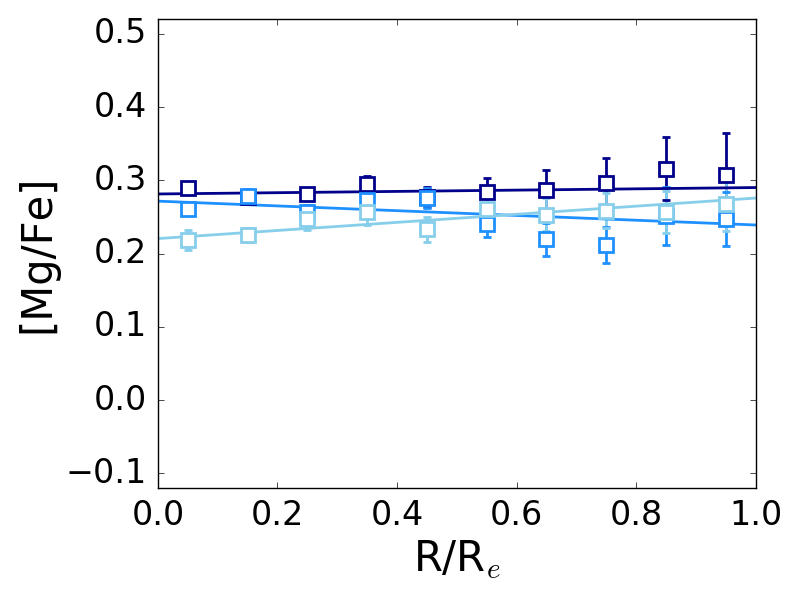}
  \includegraphics[width=.32\linewidth]{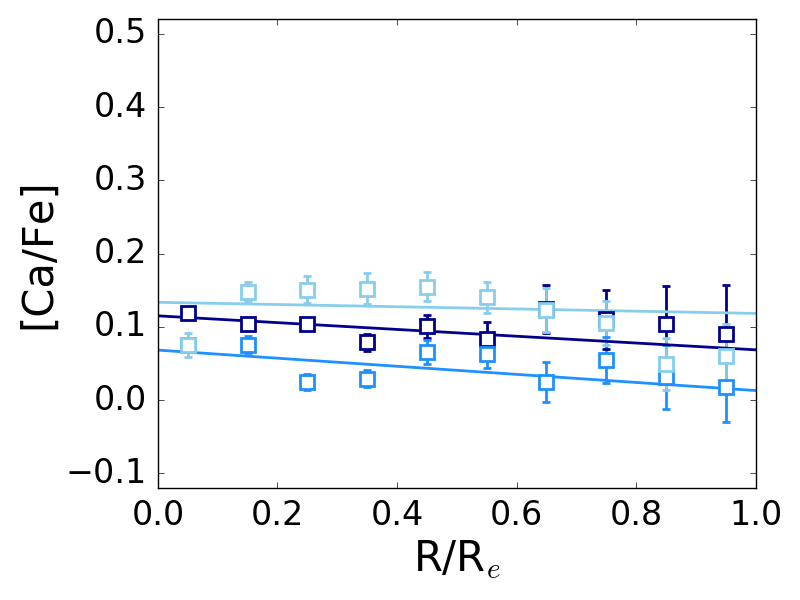}
  \includegraphics[width=.32\linewidth]{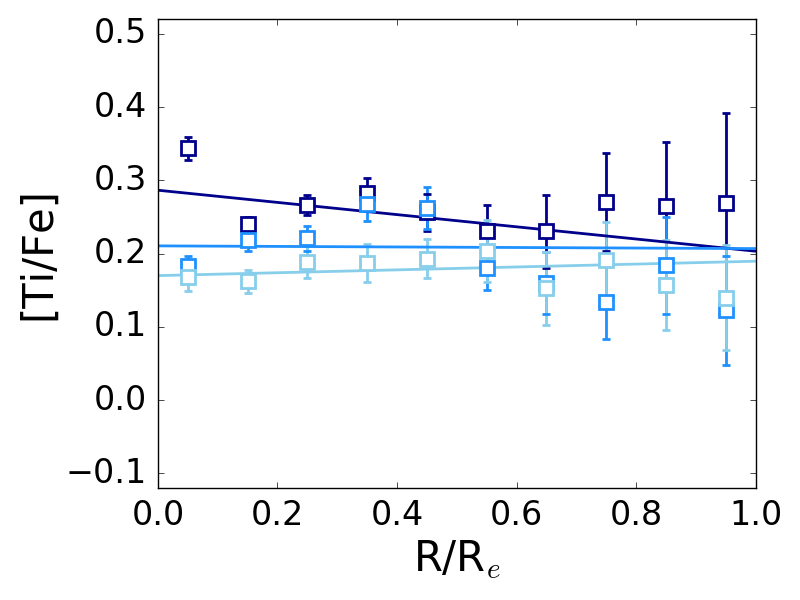}
\caption{Age and metallicity (top row) and the element abundances, C, N, Na, Mg, Ca, and Ti as a function of radius for the three mass bins. The parameters derived at each radius are shown as open squares, lines are straight line fits. The age and metallicity are fixed to the values shown in in the top panels, then the abundances shown here are varied to reproduce carefully selected indices, with [Ca/Fe] and [N/Fe] varied in separate steps as Ca4227 and CN1 are affected by other parameters (see text for more details).}
\label{fig:parameters}
\end{figure*}

Note that although early-type galaxies suffer from minimal ISM contamination, our measurements of NaD might be affected. We consider the latter negligible due to the smooth change in the NaD spectral feature as a function of radius and the small extinction values derived through full spectrum fitting as discussed in detail in \citetalias{Parikh2018}.

\subsection{Stellar population gradients}
In the following we present the radial gradients of the stellar population parameters resulting from fitting the index profiles of Fig.~\ref{fig:indices}.

\subsubsection{Age and [Z/H]}
\label{sec:age}
Stellar population age and metallicity are shown as a function of radius in the top panels of Fig.~\ref{fig:parameters}. Open squares show the derived parameter at each radial bin with 1-$\sigma$ errors, and the lines are linear fits to the profiles. Light to dark shades of blue represent increasing galaxy mass. The error on the parameter is calculated using index errors and the minimisation result; the error on the best-fit line depends on this and also on the scatter of the points about the line. These parameters are similar to \citetalias[][Fig.~11]{Parikh2018}; negligible differences arise from a better error estimation in this work, and because the fitting here does not include NaD and NaI.

Since these parameters were also derived in \citetalias{Parikh2018} and are not the focus of this work, we only briefly summarise the results, which are consistent with literature. We find that more massive galaxies are older and more metal rich \citep[e.g.][]{Kuntschner2000,Thomas2005,Thomas2010}. There is no age gradient within galaxies, but the metallicity has a steep negative gradient, which is steeper for more massive galaxies \citep{Mehlert2003, Goddard2017}.

\begin{table*}
	\centering
	\caption{Radial gradients of parameters in dex/$R_e$; age is in log Gyr/$R_e$.}
	\label{tab:gradients}
	\resizebox{\linewidth}{!}{%
	\begin{tabular}{cccccccccc}
		\hline
		Mass bin & Age & [Z/H] & [C/Fe] & [N/Fe] & [Na/Fe] & [Mg/Fe] & [Ca/Fe] & [Ti/Fe]\\
		\hline
		$9.9 - 10.2$   &		$-0.20 \pm 0.06$ & 	$-0.08 \pm 0.04$ & 	$-0.02 \pm 0.03$ & 	$ 0.12 \pm 0.04 $ & 		$-0.15 \pm 0.03$ & 	$0.06 \pm 0.01$ & 	$-0.01 \pm 0.05$ & 	$0.02 \pm 0.02$\\
		$10.2 - 10.5$ &		$-0.15 \pm 0.04$ & 	$-0.10 \pm 0.03$ & 	$-0.06 \pm 0.02$ & 	$ -0.25 \pm 0.05 $ & 		$-0.26 \pm 0.03$ & 	$-0.03 \pm 0.02$ & 	$-0.06 \pm 0.03$ & 	$0.00 \pm 0.06$\\
		$10.5 - 10.8$ &		$-0.12 \pm 0.04$ & 	$-0.11 \pm 0.03$ & 	$-0.05 \pm 0.03$ & 	$ -0.18 \pm 0.05 $ & 		$-0.29 \pm 0.02$ & 	$0.01 \pm 0.02$ & 	$-0.05 \pm 0.02$ & 	$-0.08 \pm 0.08$\\
		\hline
	\end{tabular}}
\end{table*}
\subsubsection{Element abundances}
The derived element abundances, C, N, Na, Mg, Ca, and Ti, are shown in the middle and bottom rows of Fig.~\ref{fig:parameters}. [X/Fe] is plotted as a function of radius, with the linear best fits shown as solid lines. The elements are ordered in increasing atomic number. The gradients for each mass bin are provided in Table~\ref{tab:gradients} with errors. Our method implies that all the abundances measured are also dependent on the derived basic parameters age, metallicity, and [Mg/Fe]. On top of this, [N/Fe] is measured through an index, CN1, which is also sensitive to C abundance, hence the derived N abundance is also linked to [C/Fe]. [Ca/Fe] is determined using Ca4227, which is sensitive to both C and N abundances, hence this parameter is linked to both [C/Fe] and [N/Fe].

\begin{figure*}
  \flushleft
  \includegraphics[width=.32\linewidth]{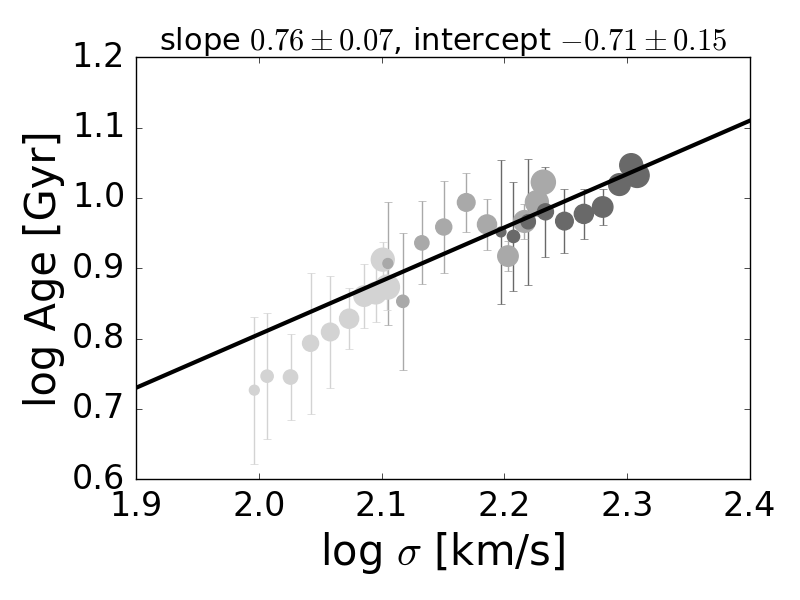}
  \includegraphics[width=.32\linewidth]{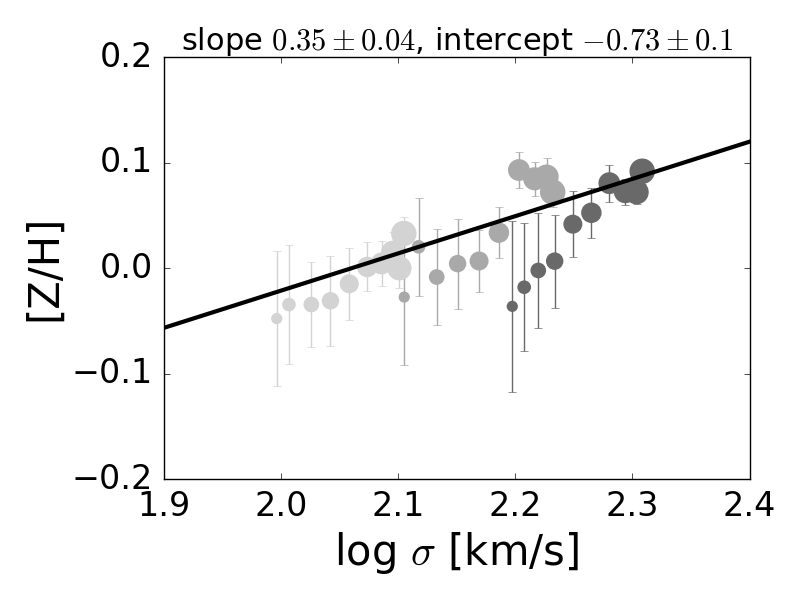}\\
  \includegraphics[width=.32\linewidth]{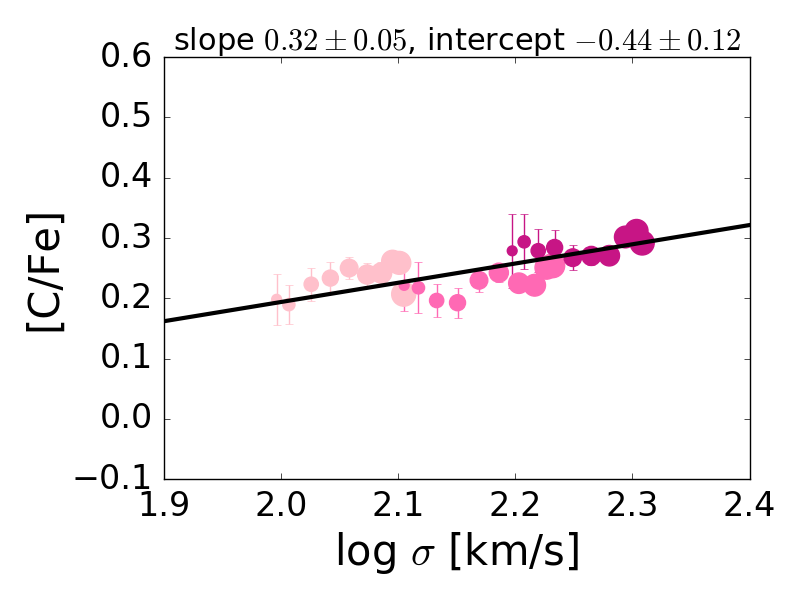}
  \includegraphics[width=.32\linewidth]{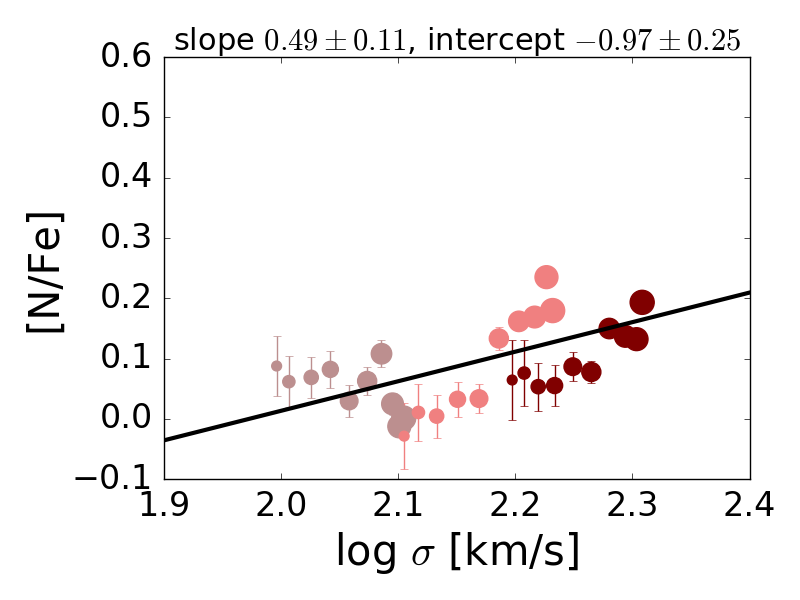}
  \includegraphics[width=.32\linewidth]{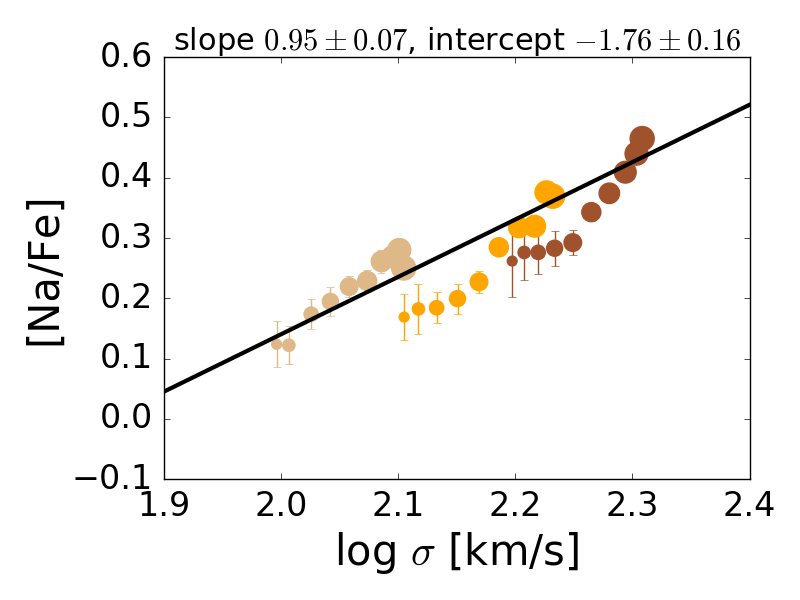}
  \includegraphics[width=.32\linewidth]{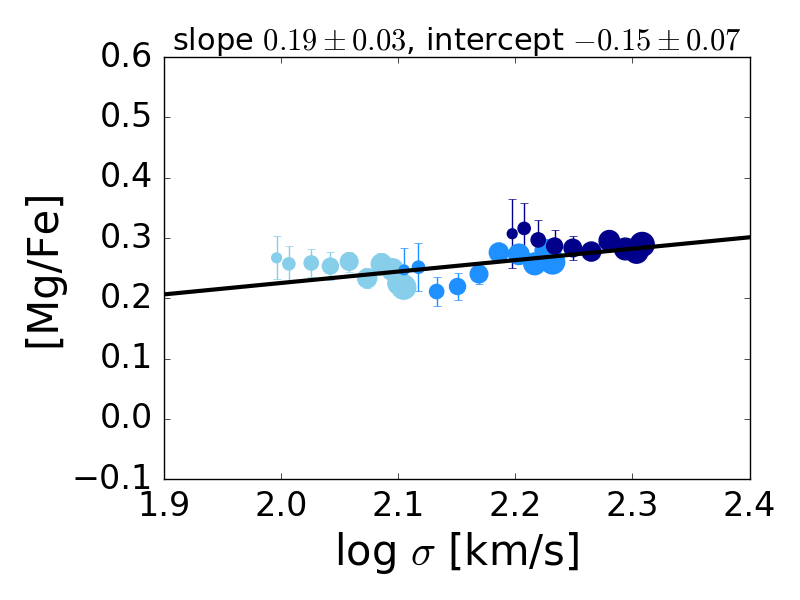}
  \includegraphics[width=.32\linewidth]{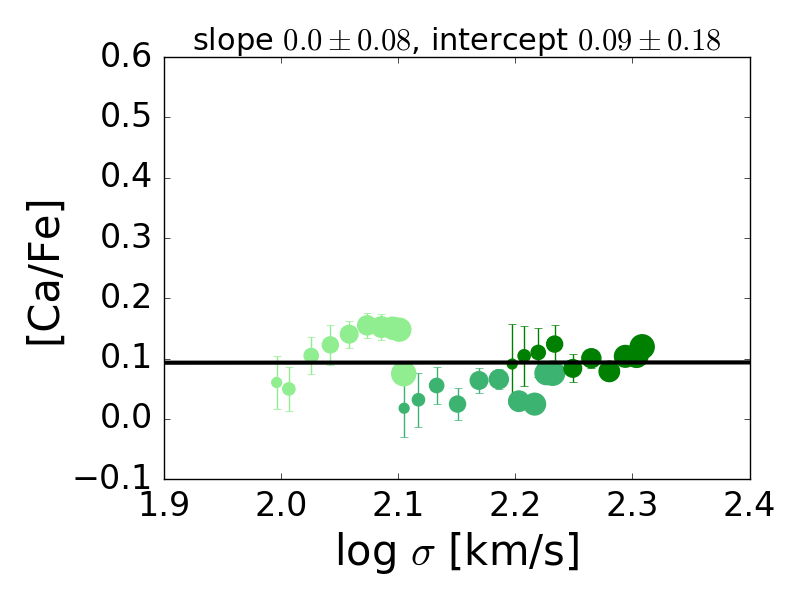}
  \includegraphics[width=.32\linewidth]{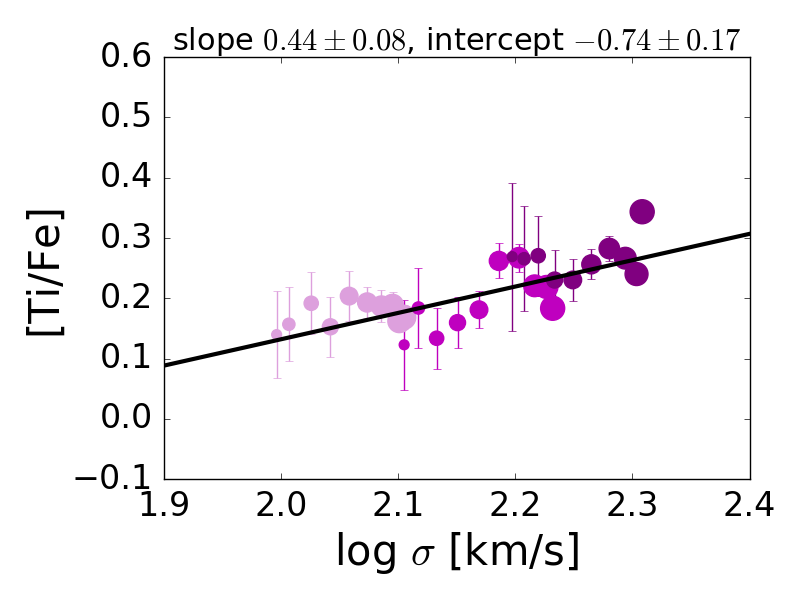}
\caption{Age, metallicity and element abundances C, N, Na, Mg, Ca, Ti plotted against local velocity dispersion as circles with 1-$\sigma$ error bars. Galaxy mass bins are shown in different shades to make them distinguishable, and decreasing circle size represents increasing radius from the centre to 1$R_e$. A linear fit is plotted in each panel, with the slope and intercept of the line given at the top.}
\label{fig:XFe}
\end{figure*}

Generally, more massive galaxies are more enhanced in all elements. [C/Fe] and [Mg/Fe] show the smallest variation with mass, from values of 0.2~dex for low mass galaxies to 0.3 for the higher mass ones. [Ti/Fe] shows more variation from 0.1 to 0.4~dex. [Ca/Fe] and [N/Fe] values all lie at or below 0.2~dex, and are close to solar values at large radii. We find maximum [Na/Fe] = +0.5~dex at the centre of the highest mass galaxies in our sample. \citetalias{Parikh2018} presents a comprehensive discussion about the implications and plausibility of the extreme [Na/Fe] abundances. The behaviour of [Ca/Fe] having values closer to zero, rather than other $\alpha$ elements like Mg, has been reported before as calcium under-abundance in elliptical galaxies \citep{Saglia2002, Cenarro2003, Thomas2003b}. The relation with galaxy mass is not well constrained from our data since our lowest mass bin is less under-abundant in Ca. Overall, our trends with galaxy mass qualitatively agree with results from global studies but interesting differences remain. A detailed comparison with the literature is presented in the Discussion section.

There is a clear difference between the abundances of C and N, with C matching Mg, while N being under-abundant relative to Mg. Such a discrepancy poses interesting clues on the formation processes of these elements in galaxies, but there is still no consensus about C and N abundances in the literature. \citetalias{Johansson2012} find the same discrepancy between C and N that we do, while \citet{Conroy2014} find N abundances higher than C abundances in massive galaxies. This difference might be a result of the treatment of these elements in the models of \citetalias{ThomasD2011b} and \citet{Conroy2012a}, i.e.\ whether or not C and N are locked to the enhanced group of elements.

Moving on to the trends with radius, we find that [Mg/Fe] mildly increases with radius for low mass galaxies. However, since the higher mass galaxies shown no gradients within $\sim 2~\sigma$, we conclude that the [Mg/Fe] radial profile is flat. This is in agreement with the the literature \citep{Mehlert2003,Kuntschner2010,Greene2015, Alton2018}, except \citet{vanDokkum2016} who obtain positive [Mg/Fe] radial gradients. All three galaxy mass bins display mildly negative [C/Fe] gradients of $\sim-0.05~dex/R_e$. \citet{Greene2015} find steeper gradients, their mass bin comparable to our sample has a [C/Fe] gradient of $-0.1~dex/R_e$. \citet{vanDokkum2016} report roughly flat [C/Fe] gradients with abundance values between 0.1 and 0.25 for galaxies of different masses. [Ca/Fe] shows similarly mild negative gradients; again consistent with the literature where [Ca/Fe] gradients appear to be roughly flat and the abundances from the different works are all consistently between 0 and 0.2~dex \citep{Greene2015, vanDokkum2016, Alton2018}.

Likewise, [Ti/Fe] shows roughly constant values with radius with no significant evidence for the presence of a gradient. However, the measurement error is large mostly because of the relatively large uncertainty in the Fe4531 index measurement on which [Ti/Fe] derivation is based (see Fig.~\ref{fig:indices}). Ti gradients are reported only in \citet{Alton2018} and they find no indication of a gradient in their stacked spectrum ($-0.02\pm0.11$). Large scatter and errors have led to global work \citepalias{Johansson2012} and local studies \citep{Alton2018} concluding that their [Ti/Fe] measurement is poorly constrained, further pointing to the difficulty in determining this particular parameter.


The only two element ratios showing significant radial gradients are [N/Fe] and [Na/Fe]. We find a clearly mass-dependent gradient in [N/Fe]. The two higher mass bins display steep negative gradients of up to $-0.25\pm 0.05$, while the low-mass galaxies in our sample, $9.9 - 10.2\;\log M_{\odot}$, have a positive gradient of $0.12\pm 0.04$. This behaviour is a direct consequence of the gradients of the CN1 index shown in Fig.~\ref{fig:indices}. This result shows that the low N abundance found in low mass galaxies by \citepalias{Johansson2012} is caused by a deficiency of N enrichment in galaxy centres rather than globally. In comparison, \citet{Greene2015} come to very different conclusions. They find [N/Fe] to be flat for all galaxy masses and their values are high ($>0.6\;$dex at all radii). Additionally, the steep gradients in C found by \citet{Greene2015} suggest that their modelling of CN features attributes variations to C whereas we attribute them to N. This work is based on stellar population models of \citet{Schiavon2007} and as mentioned before, whether or not C and N are locked to the enhanced group of elements would affect the results. It should also be noted that their N results could be affected by flux calibration uncertainties in the spectrum around the CN features. \citet{McConnell2016} compare gradients in index strength measurements of two galaxies to stellar population models and qualitatively interpret [C/Fe] gradients and flat [N/Fe].

Finally, we find strong negative radial gradients in [Na/Fe] for all galaxy mass bins \citep[see also][]{LaBarbera2016, McConnell2016, vanDokkum2016, Vaughan2017, Zieleniewski2017, Alton2018}. The slope steepens with increasing galaxy mass, with a gradient of $-0.15\pm 0.03$ for the lowest mass bin in our sample, and $-0.29\pm 0.02$ at the higher masses. Again, the steep gradients in [Na/Fe] are well reflected by the gradients in the NaD index shown in Fig.~\ref{fig:indices}. Note that all the other metallic indices also display negative gradients. These are mostly accounted for by the negative metallicity gradient alone, hence no further gradient in [X/Fe] is required to reproduce the data. Our [Na/Fe] values and gradients are consistent with the comparable mass galaxies in \citet{vanDokkum2016}. The gradients are qualitatively consistent with, albeit somewhat shallower than, the gradient of $-0.37\pm0.09$ reported by \citet{Alton2018} for their stacked spectrum.

It is encouraging that we find consistency in general with previous works, particularly the values of C, Mg, and Na. The picture is still unclear regarding Ca gradients, and the difference between C and N values. It is difficult to disentangle whether the disparity between literature stems from the use of different stellar population models or different methods of full spectrum fitting versus absorption index fitting. The discussion section further explores this as well as the physical implications of our trends.

\subsection{Global vs local relationships}
Having studied the radial gradients of the parameters, we now turn to investigate local vs global relationships of element abundance ratios with stellar velocity dispersion. The aim is to study how the index strengths change with $\sigma$ and to separate local from global relationships.

\begin{figure*}
  \includegraphics[width=.32\linewidth]{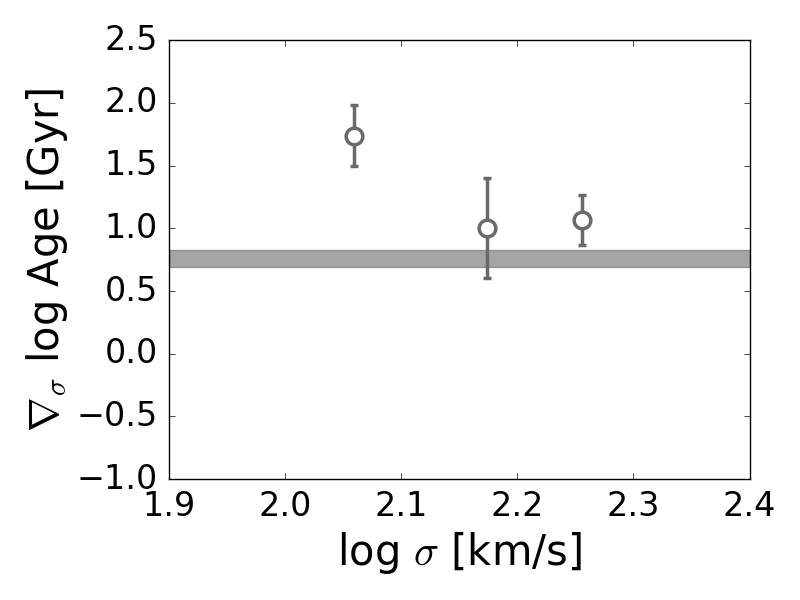}
  \includegraphics[width=.32\linewidth]{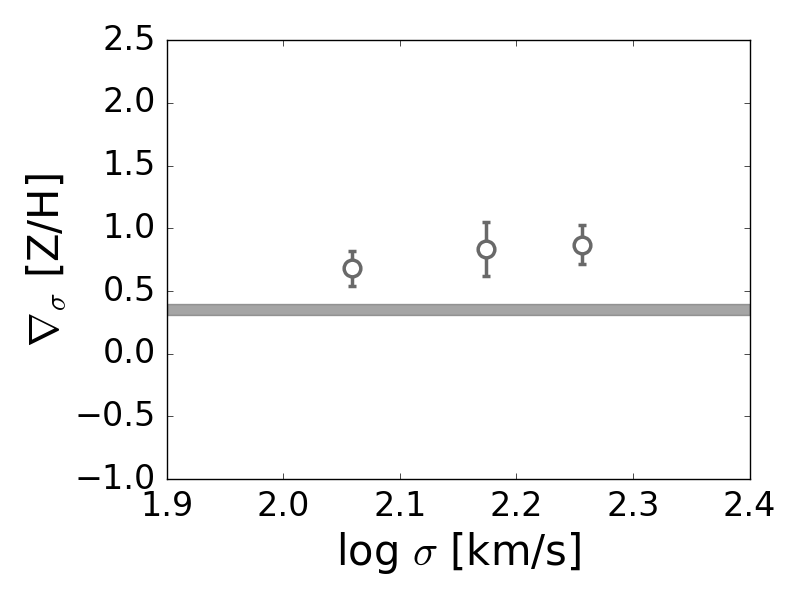}
  \includegraphics[width=.32\linewidth]{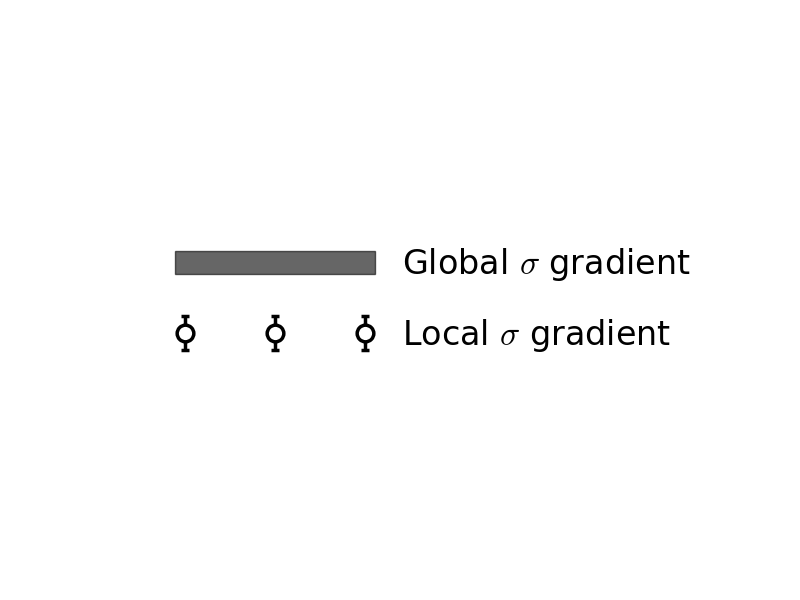}\\
  \includegraphics[width=.32\linewidth]{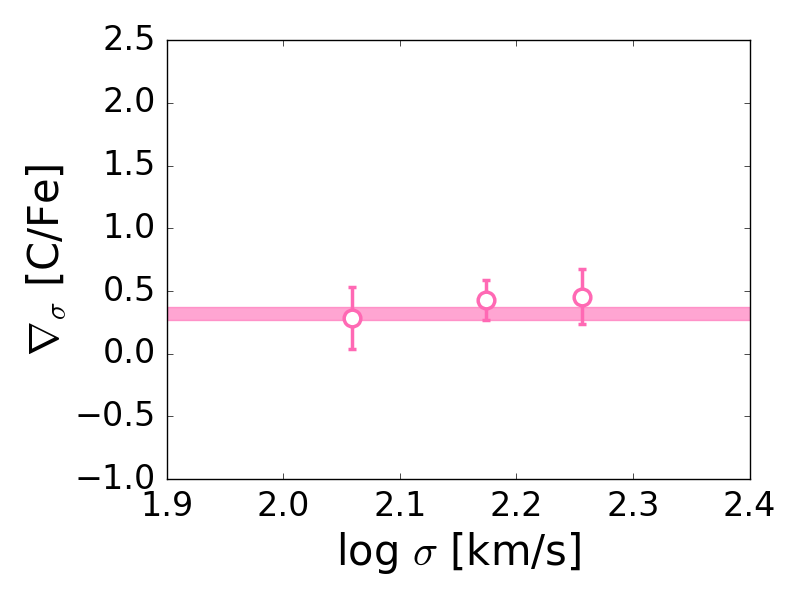}
  \includegraphics[width=.32\linewidth]{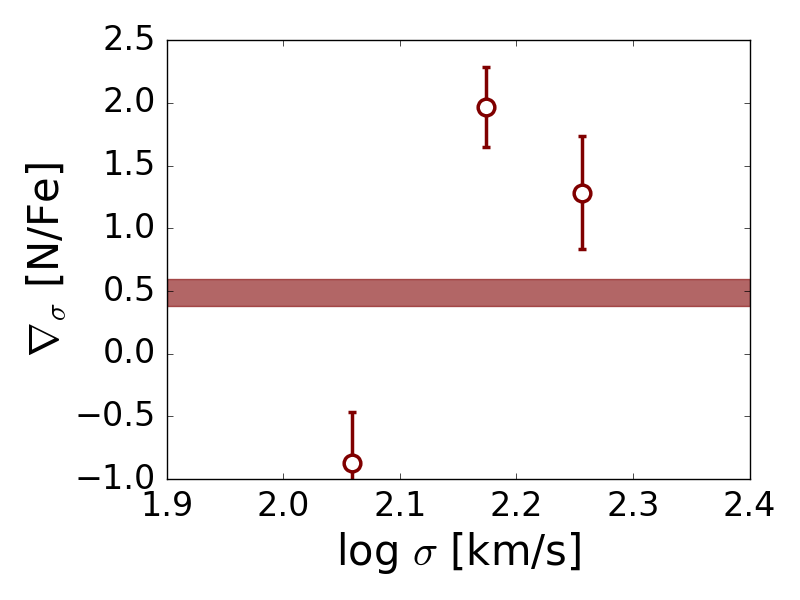}
  \includegraphics[width=.32\linewidth]{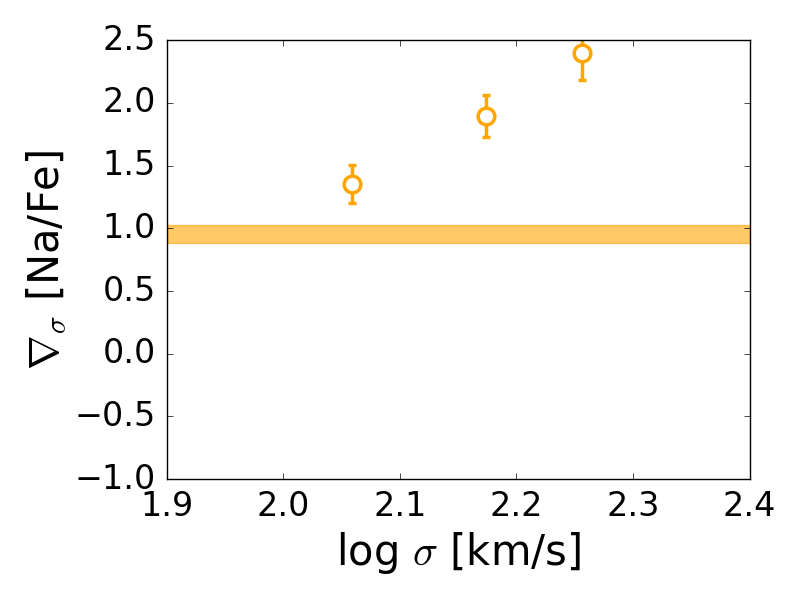}
  \includegraphics[width=.32\linewidth]{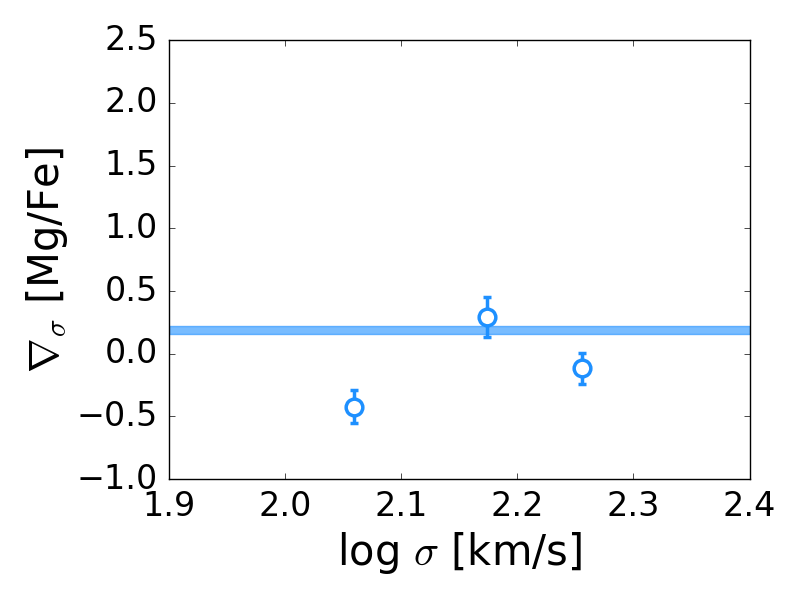}
  \includegraphics[width=.32\linewidth]{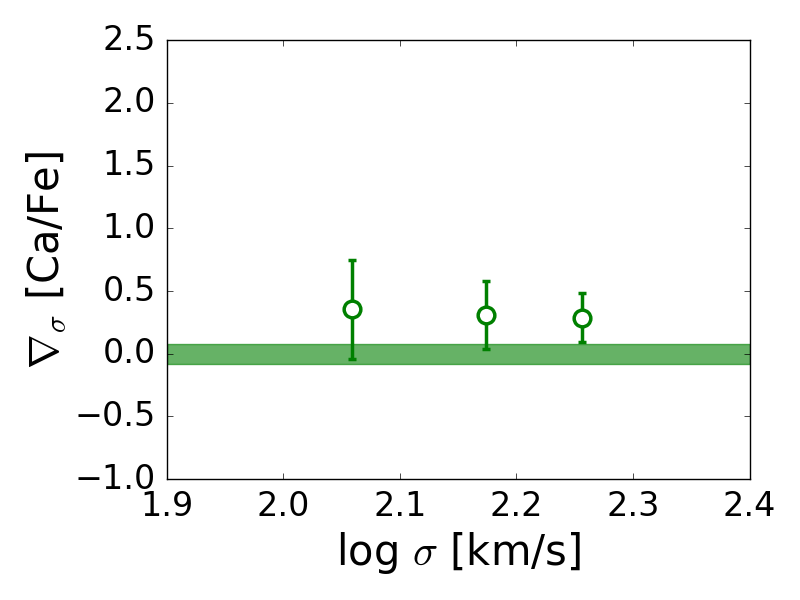}
  \includegraphics[width=.32\linewidth]{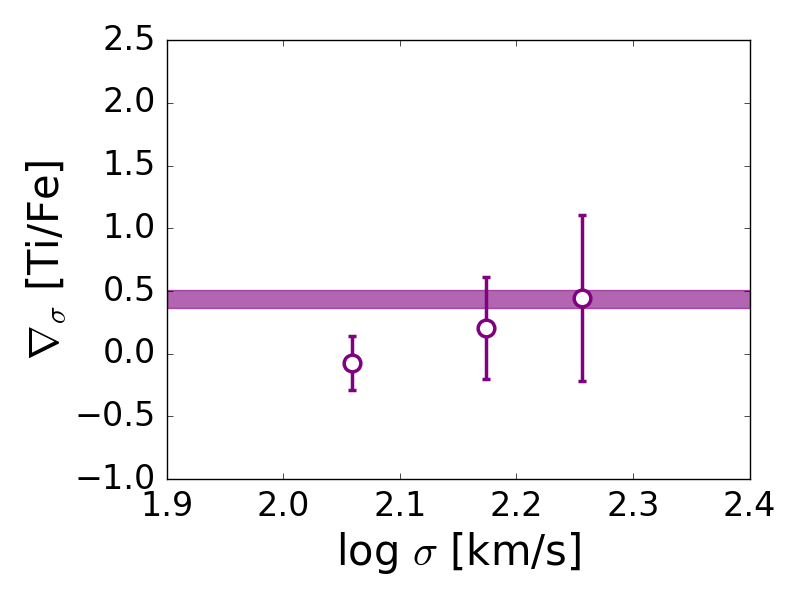}
\caption{Gradient in parameter, shaded region represents the global gradient for all data points. Circles with error bars are the gradients found within each mass bin, the x position is the mean of the $\sigma$ for that bin.}
\label{fig:gradient_comparison}
\end{figure*}

\subsubsection{Global trends with velocity dispersion}
To this end we plot the parameters versus the {\it local} velocity dispersion in Fig.~\ref{fig:XFe}. The top row shows age and metallicity, each panel thereafter shows an individual element; the circles are the derived parameters at each radius, with decreasing size representing increasing radius. The three mass bins are shown in different shades to make them distinctive and the black line is a best-fit straight line to all the data representing the {\it global} relationship. The slope and intercept of the line are given at the top of each panel. This figure allows identification of any strong radial dependence which break out from the global relationship.

As far as the global correlation between element ratio and velocity dispersion is concerned, we find that [X/Fe] increases with $\sigma$ for each element, except for Ca. Age and total metallicity also show the same behaviour and increase with $\sigma$. It is striking that [Na/Fe] shows by far the strongest correlation with $\sigma$ with a slope of $0.95\pm 0.07$.

On top of these trends, [Ca/Fe] and [N/Fe] have systematically lower values than the other element abundances for all mass bins. Overall, these relationships are consistent with the literature. We provide a detailed comparison with the literature in the Discussion section.

\subsubsection{Local trends with velocity dispersion}
Fig.~\ref{fig:XFe} allows us to assess whether there is any deviation between the local and the global dependence of the parameters with $\sigma$. Clearly the elements C and Mg, as well as stellar population age, show no separate local dependencies on top of the global relationship. The gradients with $\sigma$ are comparable for all mass bins. Instead, the elements N, Na, and total metallicity, display a striking difference between local and global $sigma$ gradients, except for the lowest mass bin. In these cases, the local dependence of the parameter on $\sigma$ is considerably steeper. Finally, the elements Ca and Ti lie between these two clear cases. There is a hint for a discrepancy between local and global dependence, but the effect appears less significant within the measurement errors.

With the aim to quantify this, we directly compare the global gradient with velocity dispersion quoted in Fig.~\ref{fig:XFe}, with the gradient calculated for each mass bin separately. This comparison is presented in Fig.~\ref{fig:gradient_comparison}. The global gradient with 1-$\sigma$ error is indicated by the shaded area, while the local gradients are the open circles with error bars placed at the mean velocity dispersion value for each mass bin.

The local gradient in stellar population age does not vary significantly across galaxies of different mass, i.e.\ the circles are roughly consistent with the shaded region. The local metallicity gradient, instead, shows a clear departure from the global relation whereby the metallicity gradients within galaxies are steeper than globally. From Fig.~\ref{fig:XFe} we identified the elements C and Mg as those where the local and global $\sigma$-dependence is consistent. This interpretation is confirmed in Fig.~\ref{fig:gradient_comparison}. The local and global $\sigma$-dependence agree perfectly for C, while there is some difference for Mg, but no clear trend with galaxy mass is evident. Also in agreement with Fig.~\ref{fig:XFe}, Ca and Ti show some systematic discrepancy between the local and the global dependence, but these are not significant within the error bars.

Fig.~\ref{fig:gradient_comparison} confirms that the two elements where the local and global dependencies are most different are N and Na. [N/Fe] is strongly $\sigma$-dependent within galaxies, and much less so globally. There appears to be a trend with galaxy mass, though. [N/Fe] is significantly enhanced at large velocity dispersion in the two higher mass-bins only.

Finally, the element Na shows the most significant signal. The $\sigma$-dependence of [Na/Fe] is considerably steeper within galaxies than globally for all three mass bins, and this discrepancy increases with increasing galaxy mass. [Na/Fe] is significantly enhanced at high velocity dispersion, particularly in massive galaxies.

\subsubsection{Element abundances relative to Mg}
\begin{figure*}
  \includegraphics[width=.32\linewidth]{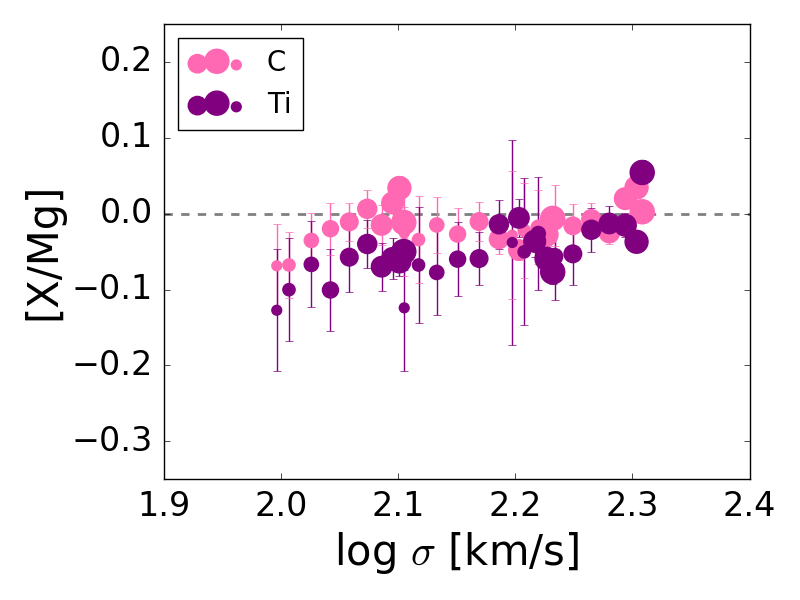}
  \includegraphics[width=.32\linewidth]{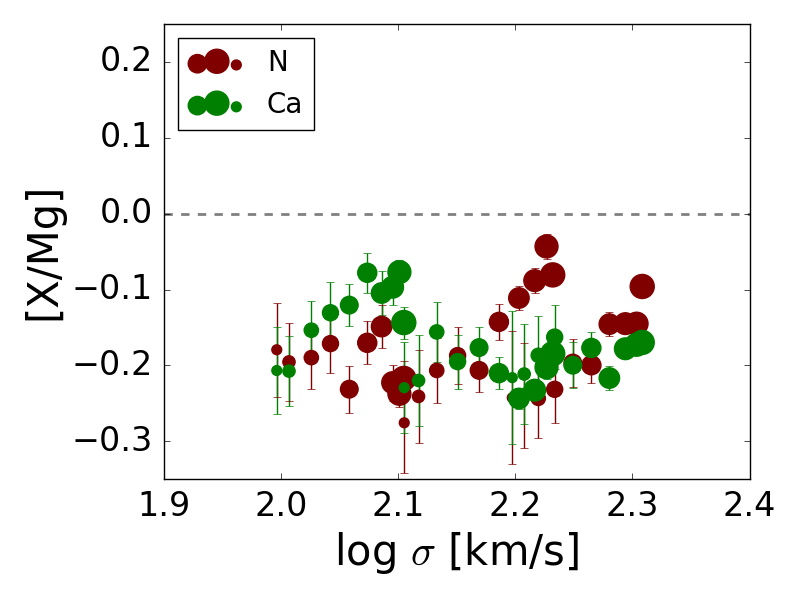}
  \includegraphics[width=.32\linewidth]{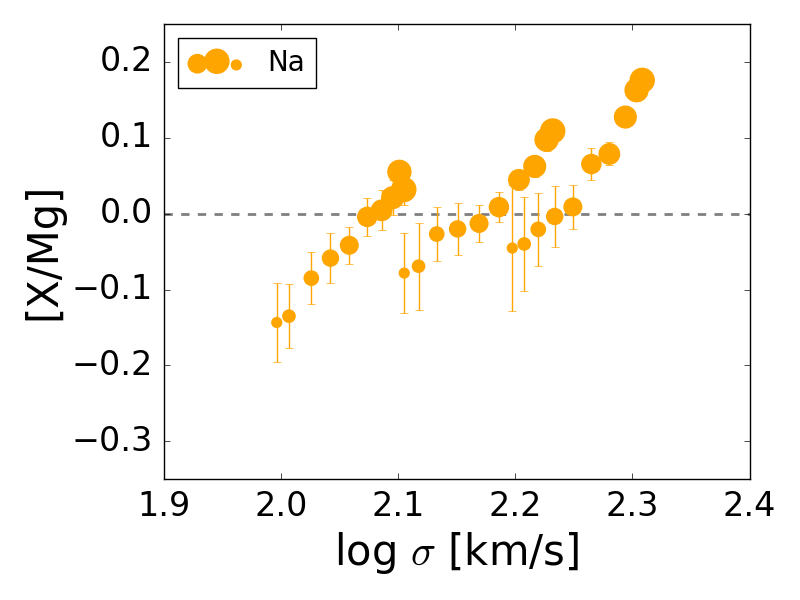}
\caption{Element abundances relative to Mg plotted against velocity dispersion for the three galaxy mass bins. Decreasing circle size represents increasing radius from the centre to 1$R_e$. Left: Elements with [X/Mg] close to zero, C and Ti, are shown. Centre: N and Ca, which are less abundant compared to Mg, are shown together. Right: Na, which is the only element strongly enhanced compared to Mg in galaxy centres.}
\label{fig:Xalpha}
\end{figure*}

Looking at the level of enhancement of the elements compared to [Mg/Fe] gives a sense of how the various elements vary. The elements can be split into 3 groups based on their [X/Mg] values: close to zero, negative or positive. These are shown as a function of velocity dispersion in Fig.~\ref{fig:Xalpha}. The dashed line at [X/Mg] = 0 indicates an abundance equivalent to Mg.

Elements which roughly belong to this category, C and Ti, are shown together in the left-hand panel of Fig.~\ref{fig:Xalpha}. It is evident that [C/Fe] (pink) closely follows [Mg/Fe], as also found in \citetalias{Johansson2012}. They find that [C/Mg] is close to zero at large $\sigma$ and -0.05 at small $\sigma$, and suggest that this could be caused by metallicity-dependent C production in massive stars. Ti (purple) is slightly under-abundant compared to C and Mg.

Elements which are strongly depleted compared to Mg are N and Ca, as shown in the middle panel. The depletion relative to Mg can be as large as 0.3~dex. Both show negative radial gradients, hence it is at large radii where these elements are the most depleted.

Finally, Na, which is the only element found to be strongly enhanced compared to Mg in galaxy centres, is shown in the right-hand panel. There is a striking dependence of [Na/Mg] with velocity dispersion, which is distinct for each mass bin, more so than [Na/Fe]. The enhancement of Na with respect to Mg goes up to +0.2~dex. \citet{LaBarbera2013} derive residual abundances for Na, Ca, and Ti, after correcting for [$\alpha$/Fe]. Our results agree well with theirs for Na and Ti, however they find residual Ca abundances to be close to zero.

\section{Discussion}
In this paper we derive radial gradients of element abundances alongside stellar ages and metallicities. We do so by measuring nine key Lick absorption indices on stacked spectra from the MaNGA survey of 366 early-type galaxies providing us with spatially resolved information out to the half-light radius. The MaNGA data allows us to explore three mass bins centred on $\log M/M_{\odot}=10, 10.4, 10.6$ corresponding to central velocity dispersions of $\sigma = 130, 170, 200\;$km/s.

In the following we compare our parameters with previous global work on abundances and also recent work using spatially resolved spectroscopy. We discuss the implications of our key results in the context of galaxy evolution and the different production pathways of the elements.

\begin{figure*}
  \includegraphics[width=.32\linewidth]{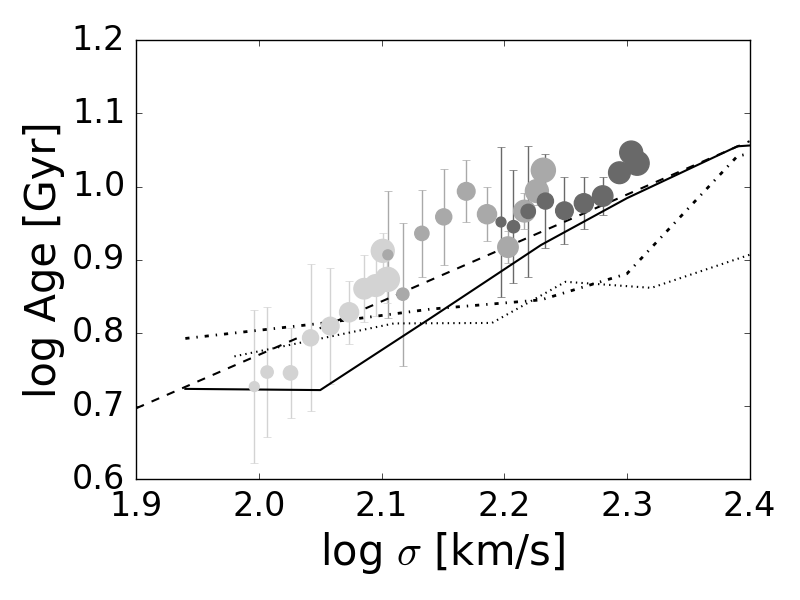}
  \includegraphics[width=.32\linewidth]{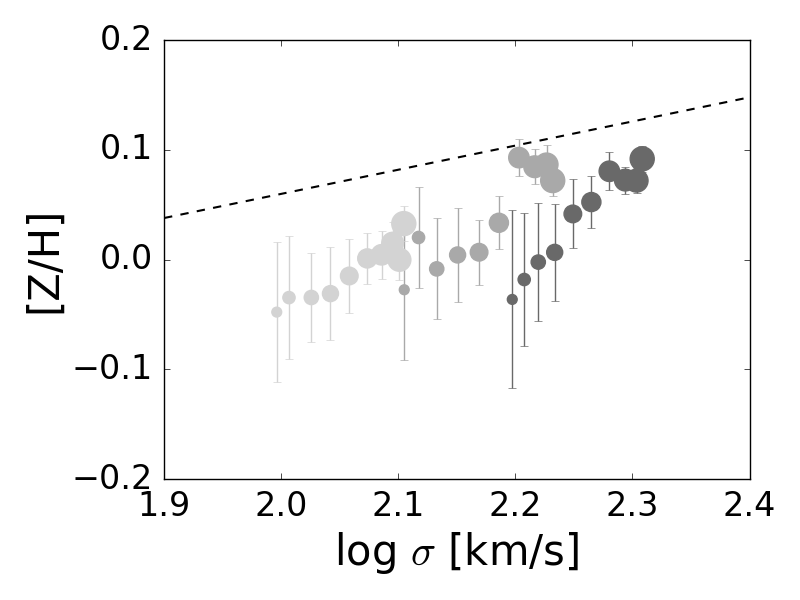}
   \includegraphics[width=.32\linewidth]{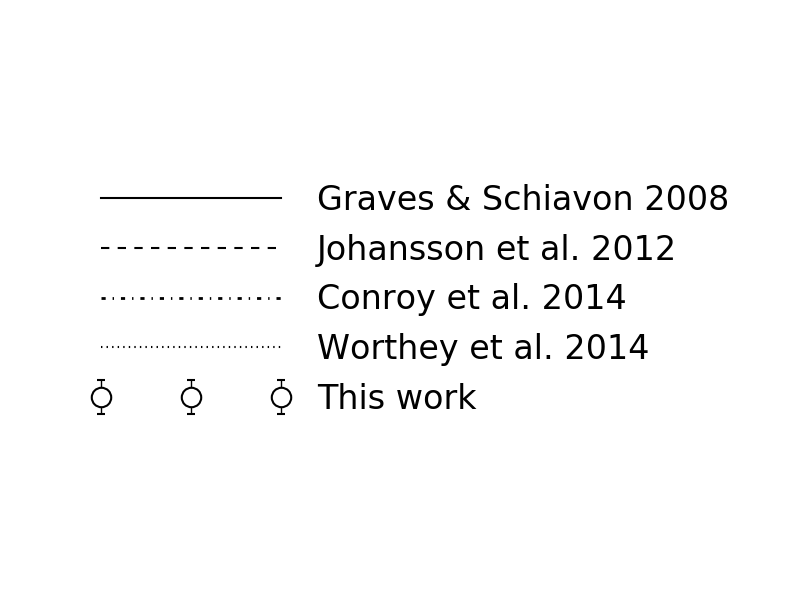}
  \includegraphics[width=.32\linewidth]{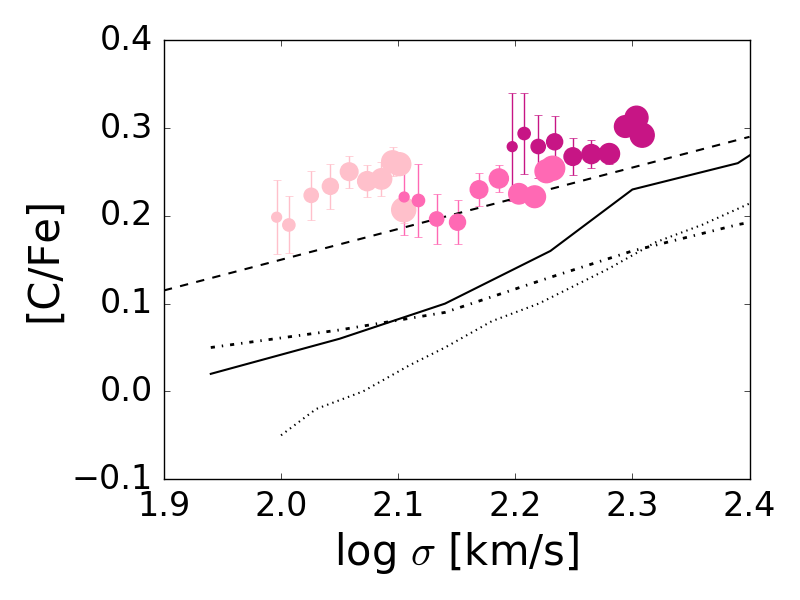}
  \includegraphics[width=.32\linewidth]{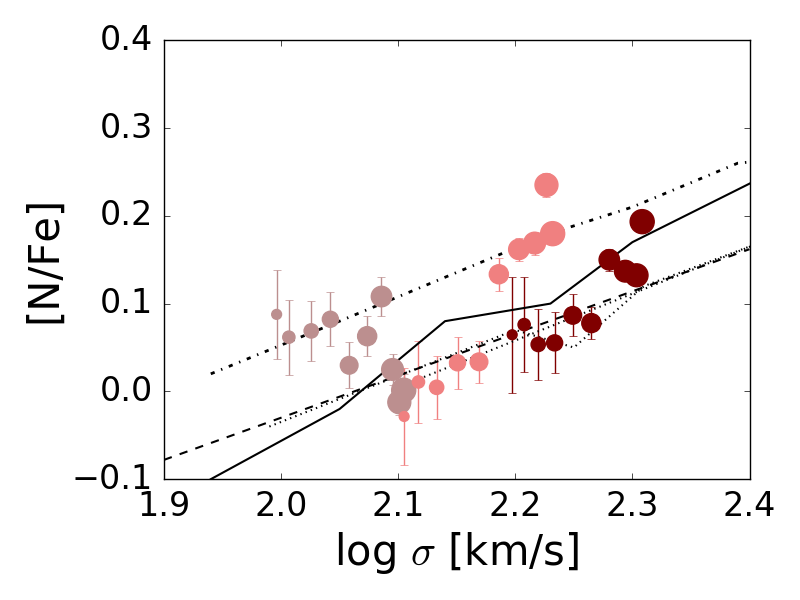}
  \includegraphics[width=.32\linewidth]{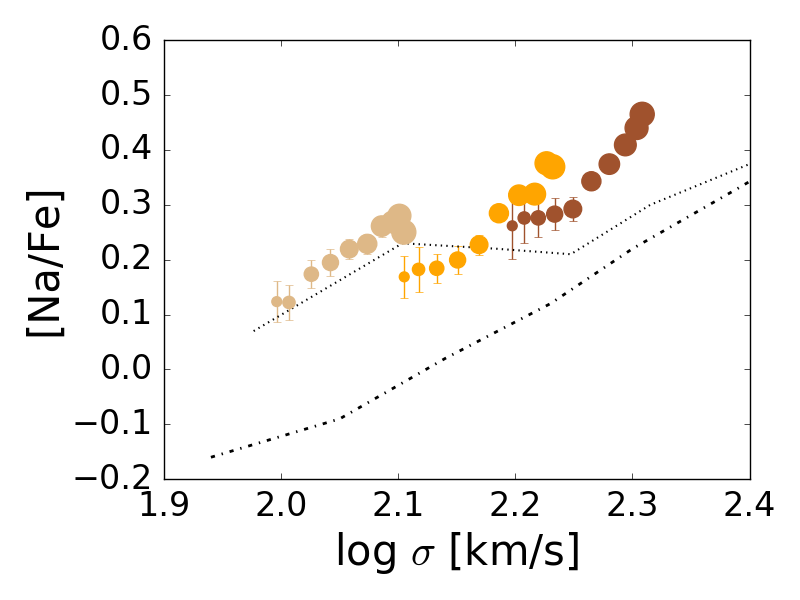}
  \includegraphics[width=.32\linewidth]{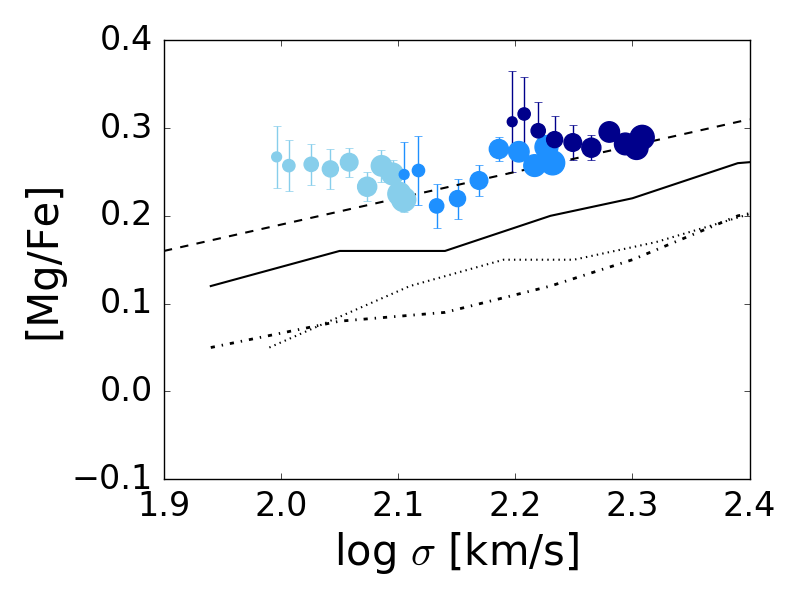}
  \includegraphics[width=.32\linewidth]{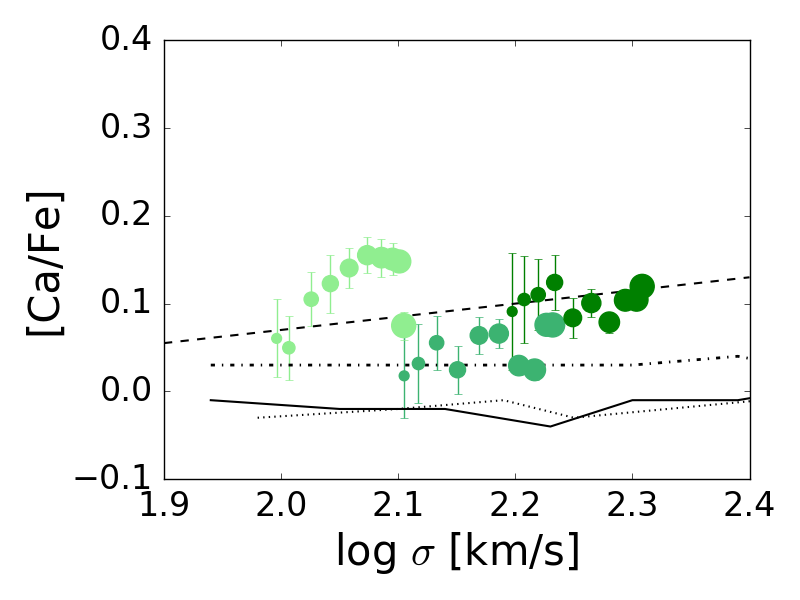}
  \includegraphics[width=.32\linewidth]{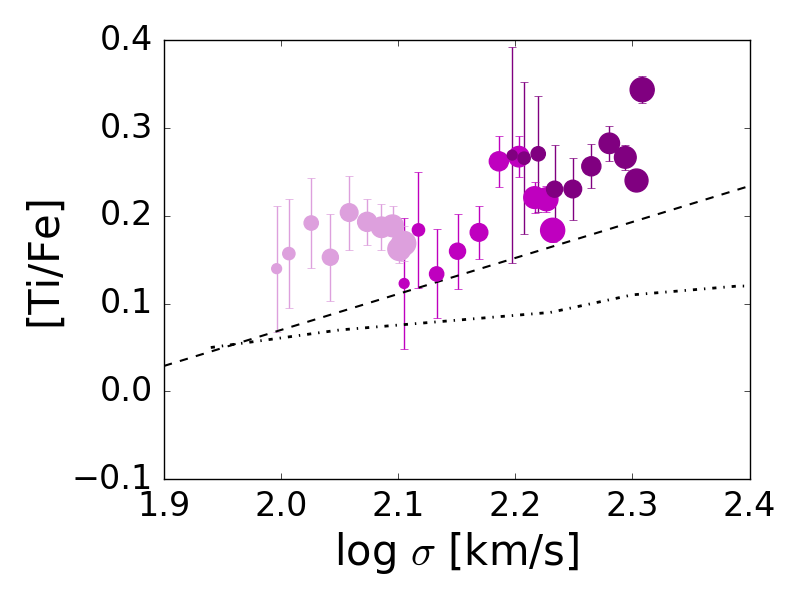}
\caption{Literature comparison: Our derived parameters are plotted as coloured circles with 1-$\sigma$ errors against velocity dispersion. As before, the three mass bins are shown in different shades, and decreasing circle size represents increasing radius. Results from \citet{Graves2008} are shown as solid lines; \citet{Johansson2012} as dashed lines; \citet{Conroy2014} as dashdot lines; \citet{Worthey2014} as dotted lines.}
\label{fig:literature_comparison}
\end{figure*}
\begin{figure*}
  \includegraphics[width=\linewidth]{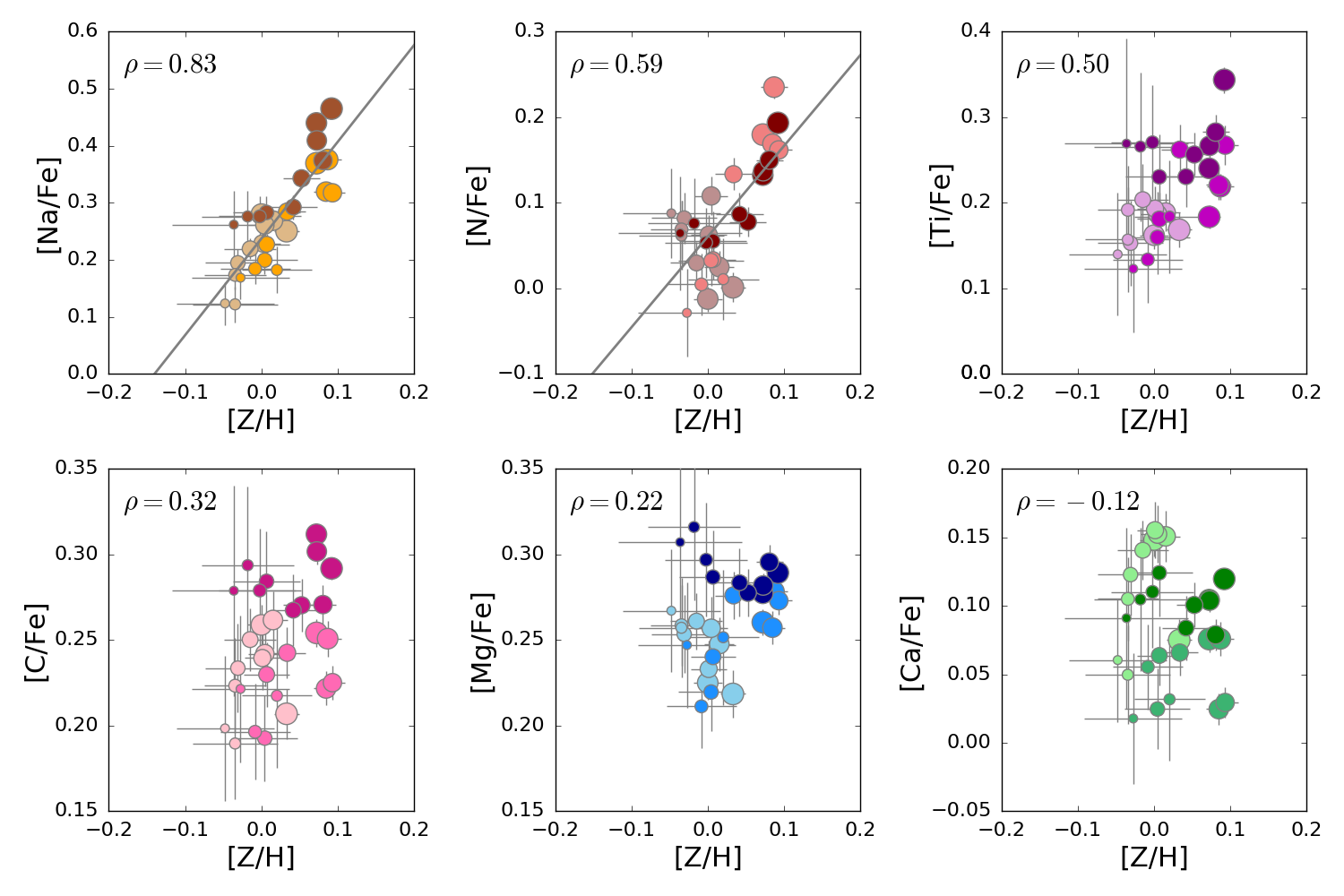}
\caption{Derived element abundances plotted against total metallicity. The mass bins are shown in different colours, decreasing circle size represents increasing radius. A linear fit to all the points is plotted if a correlation is found and the Spearman correlation coefficient is shown in each panel.}
\label{fig:correlation}
\end{figure*}
\subsection{Comparison to previous work}
A detailed comparison of our derived parameters is shown in Fig.~\ref{fig:literature_comparison}. We plot each parameter individually as a function of $\sigma$, equivalent to Fig.~\ref{fig:gradient_comparison}. The three mass bins are shown in different shades and decreasing circle size represents increasing radius.

A comprehensive comparison of stellar population parameters is presented in \citet{Conroy2014}. We expand this discussion by including the new MaNGA results presented here. Global trends from \citet{Graves2008, Johansson2012, Conroy2014, Worthey2014} are shown as solid, dashed, dashdot, and dotted black lines respectively. \citet{Graves2008} use 7 Lick indices measured on stacked spectra of SDSS galaxies and the models of \citet{Graves2007}. \citet{Conroy2014} carry out full spectrum fitting for the same spectra. These spectra, in seven bins of velocity dispersion, with mean values ranging from 88 to 300 km/s, sample the inner regions of galaxies, and hence should be compared to our innermost radial bins. \citet{Worthey2014} also use similar stacked spectra of early-type galaxies. In each of these cases, early-type galaxies were selected based on emission line cuts. Finally, \citetalias{Johansson2012} derive abundances for morphologically selected early-type galaxies from SDSS via absorption index fitting of 19 indices and the \citetalias{ThomasD2011b} models.

There is a general agreement in the literature such that all [X/Fe] abundance ratios are positively correlated with velocity dispersion. However, there is still some disagreement on the detail of these correlations and on the relative abundances of some elements. Offsets are relatively small, though, and never larger than $0.1\;$dex. The only exception is Na, with a discrepancy of $\sim 0.3\;$dex between the measured [Na/Fe] values. The origin of these differences is difficult to identify and will likely be a complex combination of differences in stellar population modelling, chemical element response functions, choice of absorption features or spectral wavelength range, fitting technique and underlying observational data.

Our results are consistent with the relationships presented in \citetalias{Johansson2012}, except that metallicities of this work are offset to slightly lower values by $\sim 0.05\;$dex. Most importantly, the discrepancy between C and N abundance with lower [N/Fe] ratios compared to [C/Fe] and [Mg/Fe] is reproduced in the present work. This agreement ought to be expected, as both studies are based on the stellar population models by \citet{Thomas2011a}.

A key result of our study echoing \citetalias{Johansson2012} is that the element C behaves similar to Mg. This finding deviates from other studies in the literature, with \citet{Graves2008}, \citet{Worthey2014}, and \citet{Conroy2014} consistently producing lower C abundances by $\sim 0.1\;$dex. Likewise, also the abundance of the element N appears to be controversial. The offset to systematically lower N abundances by $\sim 0.1\;$dex derived here and in \citetalias{Johansson2012} is in good agreement with \citet{Worthey2014}, while \citet{Conroy2014} find higher N abundances more similar to C. \citet{Graves2008} derive a steeper slope for the [N/Fe]-$\sigma$ relationship, and their values lie in between these two extremes.

The [Na/Fe] abundance ratios derived here agree well with \citet{Worthey2014}, while \citet{Conroy2014} measure values that are systematically lower by $\sim 0.3\;$dex. Interestingly, also [Mg/Fe] shows quite a large scatter of $\sim 0.2\;$dex with our and \citetalias{Johansson2012} providing the highest and \citet{Conroy2014} the lowest values. Though not shown here, \citet{DeMasi2018} also obtain Mg abundances on the higher side, ranging from 0.1 to 0.3~dex for galaxies in the same mass range as ours. \citet{Graves2008} and \citet{Worthey2014} are found in between these two extremes.

The spatially resolved work of \citet{Alton2018} derives extreme Na abundances from +0.77 to -0.11~dex, however their best-fit models are unable to reproduce all NaI features. Meanwhile, \citet{Conroy2014} derive lower [Na/Fe] values by $\sim0.2\;dex$ than our results. Although not the subject of this paper, the derived [Na/Fe] has large consequences on the derived low mass IMF slope. The abundances from the NaD index set upper limits on [Na/Fe], since this index is sensitive to Na abundance but not to the IMF; these values are then used to constrain the IMF using the NaI index which is sensitive to both Na abundance and the IMF. The higher the abundance derived, the smaller the IMF change is required. The abundances shown here explain why \citet{Alton2018} find no change in the IMF; we find modest changes between a Kroupa and a Salpeter IMF in \citetalias{Parikh2018}; the lowest abundances are derived for \citet{Conroy2014}, and hence they find steeper IMFs \citep{Conroy2012b, vanDokkum2016}. See also \citetalias[][figures 13 \& 14]{Parikh2018}, where we apply different models on the same data using the same method and derive lower or higher abundances leading to higher or lower IMFs respectively.

Finally, there appears to be better agreement on the general level of [Ca/Fe] and [Ti/Fe] enhancement. All four groups agree that [Ca/Fe] is low and close to the solar value. There is a scatter of $\sim 0.1\;$dex between the various studies with our work and \citetalias{Johansson2012} measuring the highest, and \citet{Conroy2014} measuring the lowest values. Finally, Ti is the only element where there appears to be a disagreement about the exact slope of the correlation with $\sigma$. This work and \citetalias{Johansson2012} find a steeper slope than \citet{Conroy2014} leading to discrepancies at large velocity dispersion of $\sim 0.1\;$dex.

\begin{figure*}
  \includegraphics[width=.7\linewidth]{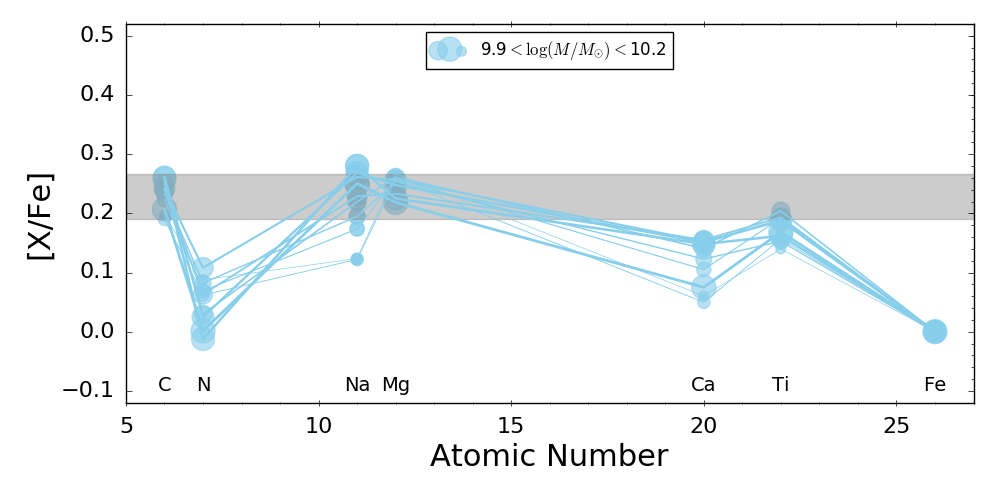}
  \includegraphics[width=.7\linewidth]{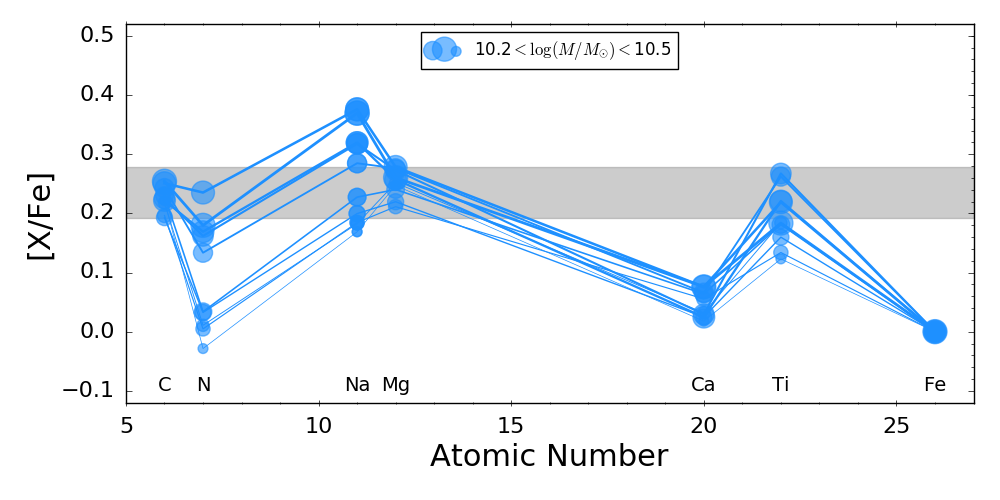}
  \includegraphics[width=.7\linewidth]{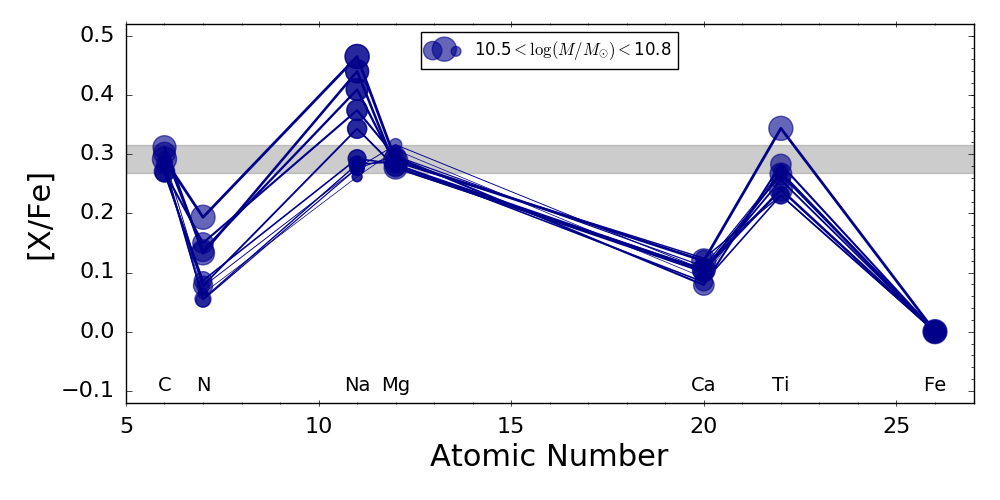}
\caption{Our results summarised in a plot similar to \citet[][Fig.~19]{Conroy2014}, but now with spatially resolved data and Na included. Element abundances are plotted as a function of atomic number. Each panel represents one of our mass bins; galaxy centres are indicated by large circles and thick lines, symbol size and line thickness decrease with increasing galaxy radius. The grey box indicates the range of [X/Fe] covered by the elements C and Mg.}
\label{fig:atomicnumber}
\end{figure*}

\subsection{Dependence on metallicity}
In order to explore the reason for the trends we see, we plot our derived abundances against the total metallicity in Fig.~\ref{fig:correlation}. As before, the mass bins are shown in different shades, and decreasing circle size represents increasing radius. The Spearman correlation coefficient is given in each panel and all the points have been fitted with a straight line when a correlation is found.

The strongest correlation with metallicity is seen for [Na/Fe]. Very importantly, this relationship is present within all three mass bins independently. This strongly suggests that the steep radial profiles as well as the steep slope of the [Na/Fe]-$\sigma$ relation are driven by metallicity-dependent Na enrichment \citep{Alton2018}. This is supported by recent nucleosynthesis calculations showing an enhancement in the production of Na in metal-rich supernovae \citep{Kobayashi2006}. Note that this result is not consistent with the calculations by \citet{Woosley1995}. 

The element N is interesting because [N/Fe] correlates well with metallicity in the two higher mass bins, although lower-mass galaxies are scattered around solar metallicity and solar [N/Fe]. The origin of N in different metallicity regimes is generally attributed to primary and secondary production channels of this element. Primary N originates from H-burning on fresh C generated by the parent star, while the secondary channel requires C and O to be originally present in the parent star and hence is only efficient at higher metallicities \citep[e.g.][]{Renzini1981,Matteucci1986}. N in early-type galaxies is expected to be entirely secondary because of the high metallicities. Our results confirm this as shown in Fig.~\ref{fig:correlation}. It will be interesting in future to extend this study to lower metallicities (larger radii and/or lower galaxy masses) to probe the plateau generally seen in dwarf galaxies at lower metallicity and N abundance \citep[e.g.,][]{Vincenzo2016}. A similar pattern is also detected in the ISM from MaNGA data \citep{Belfiore2017}.

The abundances of the remaining elements, Ti, C, Mg, and Ca, do not show any significant correlations with total metallicity. Note that while there appears to be some global correlation of [Ti/Fe] with metallicity, we do not consider it significant because of the lack thereof in the individual mass bins.

\subsection{Trends with atomic number}
In this paper we present radial gradients and correlations with velocity dispersion for the element abundance ratios [C/Fe], [N/Fe], [Na/Fe], [Mg/Fe], [Ca/Fe], and [Ti/Fe]. We have presented an assessment of links between the various element and their local vs global abundance patterns. We now summarise these results by presenting element abundance as a function of atomic number in Fig.\ref{fig:atomicnumber}. With this plot we follow and extend the approach of \citet{Conroy2014} by including spatially resolved information on top of galaxy mass. 

Fig.~\ref{fig:atomicnumber} contains three panels for the three mass bins. Galaxy centres are indicated by large circles and thick lines, symbol size and line thickness decrease with increasing galaxy radius. The grey box indicates the range of [X/Fe] covered by the elements C and Mg and also empasizes groups of elements which lie above and below these values. Many of the features discussed in this paper can be recognised in this plot. Strong radial gradients are identified by a large vertical spread of the symbols and vice versa.
\begin{itemize}
    \item C and Mg follow the same abundance pattern and display no radial gradients, as highlighted by the grey shaded regions.
    \item Ca is under-abundant closely following Fe with a negligible radial gradient in the higher mass galaxies.
    \item Ti is more enhanced than Ca, and is again more similar to the light $\alpha$-elements, C and Mg, than to Fe. There may be some gradient in the higher mass galaxies.
    \item N is depleted with respect to C and Mg, but displays a significant radial dependence.
    \item Na is enhanced, similar to C and Mg, but shows the strongest radial dependence with very high Na abundances in galaxy centres.
\end{itemize}

As discussed in \citetalias{Johansson2012}, the fact that C and Mg have similar abundance pattern sets a lower limit on the star formation time-scale, as star formation must continue over long enough periods to allow for the contribution of C from both massive and intermediate-mass stars. Here we extend the result of \citetalias{Johansson2012} to show that this must be the case at all radii. It will be interesting to test whether these element abundance gradients can be reproduced in chemical evolution simulations.

Ca has been known to be under-abundant in massive galaxies for over a decade \citep{Saglia2002,Cenarro2003,Thomas2003b}, and our results show that this feature is independent of galaxy radius. In other words, Ca under-abundance is not restricted to galaxy centres but also present in galaxy outskirts. This further supports the interpretation that Ca under-abundance in early-type galaxies is the consequence of short formation time-scales with delayed Type Ia supernovae contributing substantially to the enrichment of this element \citep{Thomas2003b}. The inference is that formation timescales do not vary much with radius, which links well with the shallow [Mg/Fe] gradients observed.

It could be expected that Ti, the next heavier $\alpha$-element after Ca, then displays the same pattern. However, this does not seem to be the case as already noticed by \citealt{Johansson2012} and \citet{Conroy2014}. It turns out that Ti enrichment appears to be again dominated by Type~II supernovae, more similar to the light $\alpha$-elements. The present results further confirm this. As opposed to Ca, Ti is again enhanced with respect to the Fe peak elements, potentially with a slight radial gradient leading to higher Ti enhancement in galaxy centres.

Nitrogen is the only other element besides Ca that reaches relatively low abundance close to Fe. In contrast to Ca, though, this is only true for the outskirts of galaxies, as there is a strong radial gradient in [N/Fe] with super-solar [N/Fe] ratios in galaxy centres. This gradient coincides with a metallicity gradient and hence is likely connected with metallicity-dependent production of secondary N. The depletion of N abundance at large radii could be caused by primordial gas in-fall suppressing secondary N production. Detailed chemical evolution models are required to explore such a scenario.

Finally, Na is the element with the strongest radial gradient leading to it being the most enhanced element in the centres of early-type galaxies. This gradient steepens with increasing galaxy mass, such that [Na/Fe] values $\sim0.2\;$dex higher than [Mg/Fe] are reached in the centre of our highest mass bin. The most plausible explanation is metallicity-dependent Na yields, but it remains to be shown with comprehensive chemical evolution simulations that such high [Na/Fe] values can indeed be reproduced with up-to-date nucleosynthesis prescriptions.

\subsection{Further Work}
The radial and mass limit in this sample comes from the number of galaxies required in each stack. As MaNGA continues to make more observations, we will be able to extend to higher mass galaxies, and possibly even to larger radii. In particular, this work does not make use of MaNGA's secondary sample which covers galaxies out to 2.5 $R_e$, owing to its lower spatial resolution than the primary+ sample. It would be interesting to study late-type galaxies and understand any differences in these parameters which might exist, especially since these are studied much less than early-type galaxies in the literature.

\section{Conclusions}
We derive chemical abundance ratios as a function of radius and galaxy mass based on data from the large galaxy IFU survey SDSS-IV/MaNGA. We analyse stacked spectra of 366 early-type galaxies adopted from our previous work in \citep{Parikh2018}. Through stacking we obtain very high S/N spectra with reduced problems due to sky residuals which allows studies out to larger radii. The bins in radius are from 0-0.1 $R_e$ to 0.9-1.0 $R_e$ in steps of 0.1 $R_e$. The resulting three mass bins, containing 122 galaxies each, are centred on $\log M/M_{\odot}$ of 10, 10.4, and 10.6, corresponding to a central velocity dispersion $\sigma$ of 130, 170, and $200\;$km/s, respectively.

We then measure absorption indices from these spectra and derive age, metallicity, and chemical element abundance ratios [X/Fe] for X = C, N, Na, Mg, Ca, and Ti. The choice of indices is based on \citetalias[][Fig.~1]{Johansson2012}. We make use of H$\beta$, Mg$b$, Fe5270, and Fe5335 as age, metallicity and [Mg/Fe] indicators. We further use C$_2$4668 and CN1 to constrain C and N, NaD to constrain Na, Ca4227 to constrain Ca, and finally Fe4531 to constrain Ti. The parameters are derived by fitting stellar population models of \citetalias{ThomasD2011b} to the data using a chi-squared minimisation code.

We recover results from the literature of no age gradient and a negative metallicity gradient that steepens with increasing galaxy mass. We further find flat or very mild gradients in [C/Fe], [Mg/Fe], [Ca/Fe], and [Ti/Fe]. The only two element ratios showing significant radial gradients are [N/Fe] and [Na/Fe]. The two higher mass bins display steep negative gradients of up to $-0.25\pm 0.05$ in [N/Fe], while the low-mass galaxies in our sample, $9.9 - 10.2\;\log M_{\odot}$, have a positive gradient of $0.12\pm 0.04$. Finally, we find strong negative radial gradients in [Na/Fe] for all galaxy mass bins. The slope steepens with increasing galaxy mass, with a gradient of $-0.15\pm 0.03$ for the lowest mass bin in our sample, and $-0.29\pm 0.02$ for the highest mass bin.

Generally we find that C traces the $\alpha$ element Mg, and the heavier $\alpha$ element Ca is under-abundant compared to C and Mg. Ti, on the other hand, is closer to C and Mg values, but is slightly less abundant at lower $\sigma$. N has similarly low values to Ca, and Na is strongly enhanced compared to Mg.

We further investigate relationships of element abundance ratios with stellar velocity dispersion with the aim to explore how the index strengths change with $\sigma$ and to separate local from global relationships. As far as the global correlation between element ratio and velocity dispersion is concerned, we find that [X/Fe] increases with $\sigma$ for each element, except for Ca. It is striking that [Na/Fe] shows by far the strongest correlation with $\sigma$ with a slope of $0.95\pm 0.05$. Interestingly, we find that the elements C and Mg show no separate local dependencies on top of the global relationship, while the elements N and Na display a striking difference between local and global $sigma$ gradients. In both cases, the local dependence of [X/Fe] on $\sigma$ is considerably steeper. The elements Ca and Ti lie between these two clear cases.

We also compare individual element abundances with the abundance of Mg. We show that C and Ti follow Mg closely, while N and Ca are depleted and Na is enhanced with respect to Mg. The element Na generally shows the most significant signal. The $\sigma$-dependence of [Na/Fe] is considerably steeper within galaxies than globally for all three mass bins, and this discrepancy increases with increasing galaxy mass. [Na/Fe] is significantly enhanced at high velocity dispersion, particularly in massive galaxies. We also test correlations of element abundance ratios with other parameters and find that [Na/Fe] is strongly correlated with metallicity in all three mass bins. This strongly suggests that the steep radial profiles as well as the steep slope of the [Na/Fe]-$\sigma$ relation are driven by metallicity-dependent Na enrichment. The only other element ratio that also correlates with metallicity is [N/Fe]. This may be attributed to primary and secondary production channels of this element.

Finally, we present the measured element abundance ratios as a function of atomic number to summarise our findings. We discuss the inferences of our results on galaxy evolution and chemical element production throughout this paper and particularly in the Discussion section. Detailed simulations of chemical enrichment are required to pin down the physical origin of our results in the framework of galaxy formation theory.

\section*{Acknowledgements}
We thank the referee, Guy Worthey, for useful insights which have improved the paper. TP is funded by a University of Portsmouth PhD bursary. The Science, Technology and Facilities Council is acknowledged for support through the Consolidated Grant Cosmology and Astrophysics at Portsmouth, ST/N000668/1. Numerical computations were performed on the Sciama High Performance Computer (HPC) cluster which is supported by the Institute of Cosmology of Gravitation, SEPnet and the University of Portsmouth. This research made use of the \textsc{python} packages \textsc{numpy} \citep{VanDerWalt2011}, \textsc{scipy} \citep{Jones2001}, \textsc{matplotlib} \citep{Hunter2007}, and \textsc{astropy} \citep{Astropy2013}.

Funding for the Sloan Digital Sky Survey IV has been provided by the Alfred P. Sloan Foundation, the U.S. Department of Energy Office of Science, and the Participating Institutions. SDSS-IV acknowledges support and resources from the Center for High-Performance Computing at the University of Utah. The SDSS web site is www.sdss.org.

SDSS-IV is managed by the Astrophysical Research Consortium for the Participating Institutions of the SDSS Collaboration including the Brazilian Participation Group, the Carnegie Institution for Science, Carnegie Mellon University, the Chilean Participation Group, the French Participation Group, Harvard-Smithsonian Center for Astrophysics, Instituto de Astrof\'isica de Canarias, The Johns Hopkins University, Kavli Institute for the Physics and Mathematics of the Universe (IPMU) / University of Tokyo, Lawrence Berkeley National Laboratory, Leibniz Institut f\"ur Astrophysik Potsdam (AIP), Max-Planck-Institut f\"ur Astronomie (MPIA Heidelberg), Max-Planck-Institut f\"ur Astrophysik (MPA Garching), Max-Planck-Institut f\"ur Extraterrestrische Physik (MPE), National Astronomical Observatories of China, New Mexico State University, New York University, University of Notre Dame, Observat\'ario Nacional / MCTI, The Ohio State University, Pennsylvania State University, Shanghai Astronomical Observatory, United Kingdom Participation Group, Universidad Nacional Aut\'onoma de M\'exico, University of Arizona, University of Colorado Boulder, University of Oxford, University of Portsmouth, University of Utah, University of Virginia, University of Washington, University of Wisconsin, Vanderbilt University, and Yale University.


\bibliographystyle{mnras}
\bibliography{IMF}

\begin{thebibliography}{}
\makeatletter
\relax
\def\mn@urlcharsother{\let\do\@makeother \do\$\do\&\do\#\do\^\do\_\do\%\do\~}
\def\mn@doi{\begingroup\mn@urlcharsother \@ifnextchar [ {\mn@doi@}
  {\mn@doi@[]}}
\def\mn@doi@[#1]#2{\def\@tempa{#1}\ifx\@tempa\@empty \href
  {http://dx.doi.org/#2} {doi:#2}\else \href {http://dx.doi.org/#2} {#1}\fi
  \endgroup}
\def\mn@eprint#1#2{\mn@eprint@#1:#2::\@nil}
\def\mn@eprint@arXiv#1{\href {http://arxiv.org/abs/#1} {{\tt arXiv:#1}}}
\def\mn@eprint@dblp#1{\href {http://dblp.uni-trier.de/rec/bibtex/#1.xml}
  {dblp:#1}}
\def\mn@eprint@#1:#2:#3:#4\@nil{\def\@tempa {#1}\def\@tempb {#2}\def\@tempc
  {#3}\ifx \@tempc \@empty \let \@tempc \@tempb \let \@tempb \@tempa \fi \ifx
  \@tempb \@empty \def\@tempb {arXiv}\fi \@ifundefined
  {mn@eprint@\@tempb}{\@tempb:\@tempc}{\expandafter \expandafter \csname
  mn@eprint@\@tempb\endcsname \expandafter{\@tempc}}}

\bibitem[\protect\citeauthoryear{{Abolfathi} et~al.,}{{Abolfathi}
  et~al.}{2017}]{Abolfathi2017}
{Abolfathi} B.,  et~al., 2017, ApJS, \href
  {http://adsabs.harvard.edu/abs/2017arXiv170709322A} {}

\bibitem[\protect\citeauthoryear{{Alton}, {Smith}  \& {Lucey}}{{Alton}
  et~al.}{2018}]{Alton2018}
{Alton} P.~D.,  {Smith} R.~J.,   {Lucey} J.~R.,  2018, \mn@doi [\mnras]
  {10.1093/mnras/sty1242}, \href
  {http://adsabs.harvard.edu/abs/2018MNRAS.tmp.1180A} {}

\bibitem[\protect\citeauthoryear{{Astropy Collaboration} et~al.,}{{Astropy
  Collaboration} et~al.}{2013}]{Astropy2013}
{Astropy Collaboration} et~al., 2013, \mn@doi [\aap]
  {10.1051/0004-6361/201322068}, \href
  {http://adsabs.harvard.edu/abs/2013A{\%}26A...558A..33A} {558, A33}

\bibitem[\protect\citeauthoryear{{Beifiori}, {Maraston}, {Thomas}  \&
  {Johansson}}{{Beifiori} et~al.}{2011}]{Beifiori2011}
{Beifiori} A.,  {Maraston} C.,  {Thomas} D.,   {Johansson} J.,  2011, \mn@doi
  [\aap] {10.1051/0004-6361/201016323}, \href
  {http://adsabs.harvard.edu/abs/2011A{\%}26A...531A.109B} {531, A109}

\bibitem[\protect\citeauthoryear{{Belfiore} et~al.,}{{Belfiore}
  et~al.}{2017}]{Belfiore2017}
{Belfiore} F.,  et~al., 2017, \mn@doi [\mnras] {10.1093/mnras/stx789}, \href
  {http://adsabs.harvard.edu/abs/2017MNRAS.469..151B} {469, 151}

\bibitem[\protect\citeauthoryear{{Bernardi}, {Nichol}, {Sheth}, {Miller}  \&
  {Brinkmann}}{{Bernardi} et~al.}{2006}]{Bernardi2006}
{Bernardi} M.,  {Nichol} R.~C.,  {Sheth} R.~K.,  {Miller} C.~J.,   {Brinkmann}
  J.,  2006, \mn@doi [\aj] {10.1086/499522}, \href
  {http://adsabs.harvard.edu/abs/2006AJ....131.1288B} {131, 1288}

\bibitem[\protect\citeauthoryear{Blanton et~al.,}{Blanton
  et~al.}{2005}]{Blanton2005}
Blanton M.~R.,  et~al., 2005, AsJ, 129, 2562

\bibitem[\protect\citeauthoryear{{Blanton} et~al.,}{{Blanton}
  et~al.}{2017}]{Blanton2017}
{Blanton} M.~R.,  et~al., 2017, \mn@doi [\aj] {10.3847/1538-3881/aa7567}, \href
  {http://adsabs.harvard.edu/abs/2017AJ....154...28B} {154, 28}

\bibitem[\protect\citeauthoryear{{Bundy} et~al.,}{{Bundy}
  et~al.}{2015}]{Bundy2015}
{Bundy} K.,  et~al., 2015, \mn@doi [\apj] {10.1088/0004-637X/798/1/7}, \href
  {http://adsabs.harvard.edu/abs/2015ApJ...798....7B} {798, 7}

\bibitem[\protect\citeauthoryear{{Cappellari} \& {Emsellem}}{{Cappellari} \&
  {Emsellem}}{2004}]{Cappellari2004}
{Cappellari} M.,  {Emsellem} E.,  2004, \mn@doi [\pasp] {10.1086/381875}, \href
  {http://adsabs.harvard.edu/abs/2004PASP..116..138C} {116, 138}

\bibitem[\protect\citeauthoryear{Cenarro, Gorgas, Vazdekis, Cardiel  \&
  Peletier}{Cenarro et~al.}{2003}]{Cenarro2003}
Cenarro A.~J.,  Gorgas J.,  Vazdekis A.,  Cardiel N.,   Peletier R.~F.,  2003,
  MNRAS, 339, 12

\bibitem[\protect\citeauthoryear{{Chabrier}}{{Chabrier}}{2003}]{Chabrier2003}
{Chabrier} G.,  2003, \mn@doi [\pasp] {10.1086/376392}, \href
  {http://adsabs.harvard.edu/abs/2003PASP..115..763C} {115, 763}

\bibitem[\protect\citeauthoryear{{Clemens}, {Bressan}, {Nikolic}, {Alexander},
  {Annibali}  \& {Rampazzo}}{{Clemens} et~al.}{2006}]{Clemens2006}
{Clemens} M.~S.,  {Bressan} A.,  {Nikolic} B.,  {Alexander} P.,  {Annibali} F.,
    {Rampazzo} R.,  2006, \mn@doi [\mnras] {10.1111/j.1365-2966.2006.10530.x},
  \href {http://adsabs.harvard.edu/abs/2006MNRAS.370..702C} {370, 702}

\bibitem[\protect\citeauthoryear{{Conroy}}{{Conroy}}{2013}]{Conroy2013}
{Conroy} C.,  2013, \mn@doi [\araa] {10.1146/annurev-astro-082812-141017},
  \href {http://adsabs.harvard.edu/abs/2013ARA%26A..51..393C} {51, 393}

\bibitem[\protect\citeauthoryear{{Conroy} \& {van Dokkum}}{{Conroy} \& {van
  Dokkum}}{2012a}]{Conroy2012a}
{Conroy} C.,  {van Dokkum} P.,  2012a, \mn@doi [\apj]
  {10.1088/0004-637X/747/1/69}, \href
  {http://adsabs.harvard.edu/abs/2012ApJ...747...69C} {747, 69}

\bibitem[\protect\citeauthoryear{{Conroy} \& {van Dokkum}}{{Conroy} \& {van
  Dokkum}}{2012b}]{Conroy2012b}
{Conroy} C.,  {van Dokkum} P.~G.,  2012b, \mn@doi [\apj]
  {10.1088/0004-637X/760/1/71}, \href
  {http://adsabs.harvard.edu/abs/2012ApJ...760...71C} {760, 71}

\bibitem[\protect\citeauthoryear{{Conroy}, {Graves}  \& {van Dokkum}}{{Conroy}
  et~al.}{2014}]{Conroy2014}
{Conroy} C.,  {Graves} G.~J.,   {van Dokkum} P.~G.,  2014, \mn@doi [\apj]
  {10.1088/0004-637X/780/1/33}, \href
  {http://adsabs.harvard.edu/abs/2014ApJ...780...33C} {780, 33}

\bibitem[\protect\citeauthoryear{{De Masi}, {Vincenzo}, {Matteucci}, {Rosani},
  {Barbera}, {Pasquali}  \& {Spitoni}}{{De Masi} et~al.}{2018}]{DeMasi2018}
{De Masi} C.,  {Vincenzo} F.,  {Matteucci} F.,  {Rosani} G.,  {Barbera} L.,
  {Pasquali} A.,   {Spitoni} E.,  2018, preprint, \href
  {http://adsabs.harvard.edu/abs/2018arXiv180506841D} {} (\mn@eprint {arXiv}
  {1805.06841})

\bibitem[\protect\citeauthoryear{Drory et~al.,}{Drory et~al.}{2015}]{Drory2015}
Drory N.,  et~al., 2015, AsJ, 149, 77

\bibitem[\protect\citeauthoryear{{Falc{\'o}n-Barroso},
  {S{\'a}nchez-Bl{\'a}zquez}, {Vazdekis}, {Ricciardelli}, {Cardiel}, {Cenarro},
  {Gorgas}  \& {Peletier}}{{Falc{\'o}n-Barroso}
  et~al.}{2011}]{Falcon-Barroso2011}
{Falc{\'o}n-Barroso} J.,  {S{\'a}nchez-Bl{\'a}zquez} P.,  {Vazdekis} A.,
  {Ricciardelli} E.,  {Cardiel} N.,  {Cenarro} A.~J.,  {Gorgas} J.,
  {Peletier} R.~F.,  2011, \mn@doi [\aap] {10.1051/0004-6361/201116842}, \href
  {http://adsabs.harvard.edu/abs/2011A{\%}26A...532A..95F} {532, A95}

\bibitem[\protect\citeauthoryear{{Goddard} et~al.,}{{Goddard}
  et~al.}{2017}]{Goddard2017}
{Goddard} D.,  et~al., 2017, \mn@doi [\mnras] {10.1093/mnras/stw3371}, \href
  {http://adsabs.harvard.edu/abs/2017MNRAS.466.4731G} {466, 4731}

\bibitem[\protect\citeauthoryear{{Graves} \& {Schiavon}}{{Graves} \&
  {Schiavon}}{2008}]{Graves2008}
{Graves} G.~J.,  {Schiavon} R.~P.,  2008, \mn@doi [The Astrophysical Journal
  Supplement Series] {10.1086/588097}, \href
  {https://ui.adsabs.harvard.edu/#abs/2008ApJS..177..446G} {177, 446}

\bibitem[\protect\citeauthoryear{{Graves}, {Faber}, {Schiavon}  \&
  {Yan}}{{Graves} et~al.}{2007}]{Graves2007}
{Graves} G.~J.,  {Faber} S.~M.,  {Schiavon} R.~P.,   {Yan} R.,  2007, \mn@doi
  [\apj] {10.1086/522325}, \href
  {http://adsabs.harvard.edu/abs/2007ApJ...671..243G} {671, 243}

\bibitem[\protect\citeauthoryear{{Greene}, {Janish}, {Ma}, {McConnell},
  {Blakeslee}, {Thomas}  \& {Murphy}}{{Greene} et~al.}{2015}]{Greene2015}
{Greene} J.~E.,  {Janish} R.,  {Ma} C.-P.,  {McConnell} N.~J.,  {Blakeslee}
  J.~P.,  {Thomas} J.,   {Murphy} J.~D.,  2015, \mn@doi [\apj]
  {10.1088/0004-637X/807/1/11}, \href
  {http://adsabs.harvard.edu/abs/2015ApJ...807...11G} {807, 11}

\bibitem[\protect\citeauthoryear{{Gunn} et~al.,}{{Gunn}
  et~al.}{2006}]{Gunn2006}
{Gunn} J.~E.,  et~al., 2006, \mn@doi [\aj] {10.1086/500975}, \href
  {http://adsabs.harvard.edu/abs/2006AJ....131.2332G} {131, 2332}

\bibitem[\protect\citeauthoryear{Hunter}{Hunter}{2007}]{Hunter2007}
Hunter J.~D.,  2007, \mn@doi [Computing In Science \& Engineering]
  {10.1109/MCSE.2007.55}, 9, 90

\bibitem[\protect\citeauthoryear{Johansson, Thomas  \& Maraston}{Johansson
  et~al.}{2010}]{Johansson2010}
Johansson J.,  Thomas D.,   Maraston C.,  2010, MNRAS, 19, 1

\bibitem[\protect\citeauthoryear{Johansson, Thomas  \& Maraston}{Johansson
  et~al.}{2012}]{Johansson2012}
Johansson J.,  Thomas D.,   Maraston C.,  2012, MNRAS, 421, 1908

\bibitem[\protect\citeauthoryear{Jones, Oliphant, Peterson  et~al.}{Jones
  et~al.}{2001}]{Jones2001}
Jones E.,  Oliphant T.,  Peterson P.,   et~al., 2001, {SciPy}: Open source
  scientific tools for {Python}, \url {http://www.scipy.org/}

\bibitem[\protect\citeauthoryear{{Kobayashi}, {Umeda}, {Nomoto}, {Tominaga}  \&
  {Ohkubo}}{{Kobayashi} et~al.}{2006}]{Kobayashi2006}
{Kobayashi} C.,  {Umeda} H.,  {Nomoto} K.,  {Tominaga} N.,   {Ohkubo} T.,
  2006, \mn@doi [\apj] {10.1086/508914}, \href
  {http://adsabs.harvard.edu/abs/2006ApJ...653.1145K} {653, 1145}

\bibitem[\protect\citeauthoryear{{Korn}, {Maraston}  \& {Thomas}}{{Korn}
  et~al.}{2005}]{Korn2005}
{Korn} A.~J.,  {Maraston} C.,   {Thomas} D.,  2005, \mn@doi [\aap]
  {10.1051/0004-6361:20042126}, \href
  {http://adsabs.harvard.edu/abs/2005A{\%}26A...438..685K} {438, 685}

\bibitem[\protect\citeauthoryear{Kuntschner}{Kuntschner}{2000}]{Kuntschner2000}
Kuntschner H.,  2000, Astrophys. Space Sci., 315, 184

\bibitem[\protect\citeauthoryear{{Kuntschner}}{{Kuntschner}}{2004}]{Kuntschner2010}
{Kuntschner} H.,  2004, \mn@doi [\aap] {10.1051/0004-6361:20041414}, \href
  {http://adsabs.harvard.edu/abs/2004A{\%}26A...426..737K} {426, 737}

\bibitem[\protect\citeauthoryear{{La Barbera}, {Ferreras}, {Vazdekis}, {de la
  Rosa}, {de Carvalho}, {Trevisan}, {Falc{\'o}n-Barroso}  \&
  {Ricciardelli}}{{La Barbera} et~al.}{2013}]{LaBarbera2013}
{La Barbera} F.,  {Ferreras} I.,  {Vazdekis} A.,  {de la Rosa} I.~G.,  {de
  Carvalho} R.~R.,  {Trevisan} M.,  {Falc{\'o}n-Barroso} J.,   {Ricciardelli}
  E.,  2013, \mn@doi [\mnras] {10.1093/mnras/stt943}, \href
  {http://adsabs.harvard.edu/abs/2013MNRAS.433.3017L} {433, 3017}

\bibitem[\protect\citeauthoryear{{La Barbera}, {Vazdekis}, {Ferreras},
  {Pasquali}, {Cappellari}, {Mart{\'{\i}}n-Navarro}, {Sch{\"o}nebeck}  \&
  {Falc{\'o}n-Barroso}}{{La Barbera} et~al.}{2016}]{LaBarbera2016}
{La Barbera} F.,  {Vazdekis} A.,  {Ferreras} I.,  {Pasquali} A.,  {Cappellari}
  M.,  {Mart{\'{\i}}n-Navarro} I.,  {Sch{\"o}nebeck} F.,   {Falc{\'o}n-Barroso}
  J.,  2016, \mn@doi [\mnras] {10.1093/mnras/stv2996}, \href
  {http://adsabs.harvard.edu/abs/2016MNRAS.457.1468L} {457, 1468}

\bibitem[\protect\citeauthoryear{{Law} et~al.,}{{Law} et~al.}{2015}]{Law2015}
{Law} D.~R.,  et~al., 2015, \mn@doi [\aj] {10.1088/0004-6256/150/1/19}, \href
  {http://adsabs.harvard.edu/abs/2015AJ....150...19L} {150, 19}

\bibitem[\protect\citeauthoryear{{Law} et~al.,}{{Law} et~al.}{2016}]{Law2016}
{Law} D.~R.,  et~al., 2016, \mn@doi [\aj] {10.3847/0004-6256/152/4/83}, \href
  {http://adsabs.harvard.edu/abs/2016AJ....152...83L} {152, 83}

\bibitem[\protect\citeauthoryear{{Lintott} et~al.,}{{Lintott}
  et~al.}{2011}]{Lintott2011}
{Lintott} C.,  et~al., 2011, \mn@doi [\mnras]
  {10.1111/j.1365-2966.2010.17432.x}, \href
  {http://adsabs.harvard.edu/abs/2011MNRAS.410..166L} {410, 166}

\bibitem[\protect\citeauthoryear{{Maraston}}{{Maraston}}{1998}]{Maraston1998}
{Maraston} C.,  1998, \mnras, 300, 872

\bibitem[\protect\citeauthoryear{{Maraston}}{{Maraston}}{2005}]{Maraston2005}
{Maraston} C.,  2005, \mn@doi [\mnras] {10.1111/j.1365-2966.2005.09270.x},
  \href {http://adsabs.harvard.edu/abs/2005MNRAS.362..799M} {362, 799}

\bibitem[\protect\citeauthoryear{Maraston \& Str{\"{o}}mb{\"{a}}ck}{Maraston \&
  Str{\"{o}}mb{\"{a}}ck}{2011}]{Maraston2011}
Maraston C.,  Str{\"{o}}mb{\"{a}}ck G.,  2011, MNRAS, 418, 2785

\bibitem[\protect\citeauthoryear{{Matteucci}}{{Matteucci}}{1986}]{Matteucci1986}
{Matteucci} F.,  1986, \mn@doi [\mnras] {10.1093/mnras/221.4.911}, \href
  {http://adsabs.harvard.edu/abs/1986MNRAS.221..911M} {221, 911}

\bibitem[\protect\citeauthoryear{{Matteucci}}{{Matteucci}}{1994}]{Matteucci1994}
{Matteucci} F.,  1994, \aap, \href
  {http://adsabs.harvard.edu/abs/1994A%26A...288...57M} {288, 57}

\bibitem[\protect\citeauthoryear{McConnell, Lu  \& Mann}{McConnell
  et~al.}{2016}]{McConnell2016}
McConnell N.~J.,  Lu J.~R.,   Mann A.~W.,  2016, ApJ, 821, 39

\bibitem[\protect\citeauthoryear{{Mehlert}, {Thomas}, {Saglia}, {Bender}  \&
  {Wegner}}{{Mehlert} et~al.}{2003}]{Mehlert2003}
{Mehlert} D.,  {Thomas} D.,  {Saglia} R.~P.,  {Bender} R.,   {Wegner} G.,
  2003, \mn@doi [\aap] {10.1051/0004-6361:20030886}, \href
  {http://adsabs.harvard.edu/abs/2003A{\%}26A...407..423M} {407, 423}

\bibitem[\protect\citeauthoryear{Parikh et~al.,}{Parikh
  et~al.}{2018}]{Parikh2018}
Parikh T.,  et~al., 2018, \mn@doi [\mnras] {10.1093/mnras/sty785}, 477, 3954

\bibitem[\protect\citeauthoryear{{Pipino}, {Chiappini}, {Graves}  \&
  {Matteucci}}{{Pipino} et~al.}{2009}]{Pipino2009}
{Pipino} A.,  {Chiappini} C.,  {Graves} G.,   {Matteucci} F.,  2009, \mn@doi
  [\mnras] {10.1111/j.1365-2966.2009.14833.x}, \href
  {http://adsabs.harvard.edu/abs/2009MNRAS.396.1151P} {396, 1151}

\bibitem[\protect\citeauthoryear{{Price}, {Phillipps}, {Huxor}, {Smith}  \&
  {Lucey}}{{Price} et~al.}{2011}]{Price2011}
{Price} J.,  {Phillipps} S.,  {Huxor} A.,  {Smith} R.~J.,   {Lucey} J.~R.,
  2011, \mn@doi [\mnras] {10.1111/j.1365-2966.2010.17862.x}, \href
  {http://adsabs.harvard.edu/abs/2011MNRAS.411.2558P} {411, 2558}

\bibitem[\protect\citeauthoryear{{Prugniel}, {Vauglin}  \& {Koleva}}{{Prugniel}
  et~al.}{2011}]{Prugniel2011}
{Prugniel} P.,  {Vauglin} I.,   {Koleva} M.,  2011, \aap, 531, A165

\bibitem[\protect\citeauthoryear{{Renzini} \& {Voli}}{{Renzini} \&
  {Voli}}{1981}]{Renzini1981}
{Renzini} A.,  {Voli} M.,  1981, \aap, \href
  {https://ui.adsabs.harvard.edu/#abs/1981A&A....94..175R} {500, 221}

\bibitem[\protect\citeauthoryear{Saglia, Maraston, Thomas, Bender  \&
  Colless}{Saglia et~al.}{2002}]{Saglia2002}
Saglia R.~P.,  Maraston C.,  Thomas D.,  Bender R.,   Colless M.,  2002, ApJ,
  579, L13

\bibitem[\protect\citeauthoryear{{Salpeter}}{{Salpeter}}{1955}]{Salpeter1955}
{Salpeter} E.~E.,  1955, \mn@doi [\apj] {10.1086/145971}, \href
  {http://adsabs.harvard.edu/abs/1955ApJ...121..161S} {121, 161}

\bibitem[\protect\citeauthoryear{{S{\'a}nchez-Bl{\'a}zquez}, {Gorgas},
  {Cardiel}, {Cenarro}  \& {Gonz{\'a}lez}}{{S{\'a}nchez-Bl{\'a}zquez}
  et~al.}{2003}]{Sanchez-Blazquez2003}
{S{\'a}nchez-Bl{\'a}zquez} P.,  {Gorgas} J.,  {Cardiel} N.,  {Cenarro} J.,
  {Gonz{\'a}lez} J.~J.,  2003, \mn@doi [\apjl] {10.1086/376825}, \href
  {http://adsabs.harvard.edu/abs/2003ApJ...590L..91S} {590, L91}

\bibitem[\protect\citeauthoryear{{S{\'a}nchez-Bl{\'a}zquez}
  et~al.,}{{S{\'a}nchez-Bl{\'a}zquez} et~al.}{2006}]{Sanchez-Blazquez2006}
{S{\'a}nchez-Bl{\'a}zquez} P.,  et~al., 2006, \mn@doi [\mnras]
  {10.1111/j.1365-2966.2006.10699.x}, \href
  {http://adsabs.harvard.edu/abs/2006MNRAS.371..703S} {371, 703}

\bibitem[\protect\citeauthoryear{{Schiavon}}{{Schiavon}}{2007}]{Schiavon2007}
{Schiavon} R.~P.,  2007, \mn@doi [\apjs] {10.1086/511753}, \href
  {http://adsabs.harvard.edu/abs/2007ApJS..171..146S} {171, 146}

\bibitem[\protect\citeauthoryear{{Smee} et~al.,}{{Smee}
  et~al.}{2013}]{Smee2013}
{Smee} S.~A.,  et~al., 2013, \mn@doi [\aj] {10.1088/0004-6256/146/2/32}, \href
  {http://adsabs.harvard.edu/abs/2013AJ....146...32S} {146, 32}

\bibitem[\protect\citeauthoryear{{Thomas}, {Greggio}  \& {Bender}}{{Thomas}
  et~al.}{1999}]{Thomas1999}
{Thomas} D.,  {Greggio} L.,   {Bender} R.,  1999, \mn@doi [\mnras]
  {10.1046/j.1365-8711.1999.02138.x}, \href
  {http://adsabs.harvard.edu/abs/1999MNRAS.302..537T} {302, 537}

\bibitem[\protect\citeauthoryear{{Thomas}, {Maraston}  \& {Bender}}{{Thomas}
  et~al.}{2003a}]{Thomas2003}
{Thomas} D.,  {Maraston} C.,   {Bender} R.,  2003a, \mn@doi [\mnras]
  {10.1046/j.1365-8711.2003.06248.x}, \href
  {http://adsabs.harvard.edu/abs/2003MNRAS.339..897T} {339, 897}

\bibitem[\protect\citeauthoryear{{Thomas}, {Maraston}  \& {Bender}}{{Thomas}
  et~al.}{2003b}]{Thomas2003b}
{Thomas} D.,  {Maraston} C.,   {Bender} R.,  2003b, \mn@doi [\mnras]
  {10.1046/j.1365-8711.2003.06659.x}, \href
  {https://ui.adsabs.harvard.edu/#abs/2003MNRAS.343..279T} {343, 279}

\bibitem[\protect\citeauthoryear{Thomas, Maraston  \& Korn}{Thomas
  et~al.}{2004}]{Thomas2004}
Thomas D.,  Maraston C.,   Korn A.,  2004, MNRAS, 351, L19

\bibitem[\protect\citeauthoryear{{Thomas}, {Maraston}, {Bender}  \& {Mendes de
  Oliveira}}{{Thomas} et~al.}{2005}]{Thomas2005}
{Thomas} D.,  {Maraston} C.,  {Bender} R.,   {Mendes de Oliveira} C.,  2005,
  \mn@doi [\apj] {10.1086/426932}, \href
  {http://adsabs.harvard.edu/abs/2005ApJ...621..673T} {621, 673}

\bibitem[\protect\citeauthoryear{Thomas, Maraston, Schawinski, Sarzi  \&
  Silk}{Thomas et~al.}{2010}]{Thomas2010}
Thomas D.,  Maraston C.,  Schawinski K.,  Sarzi M.,   Silk J.,  2010, MNRAS,
  404, 1775

\bibitem[\protect\citeauthoryear{Thomas, Maraston  \& Johansson}{Thomas
  et~al.}{2011a}]{ThomasD2011b}
Thomas D.,  Maraston C.,   Johansson J.,  2011a, MNRAS, 412, 2183

\bibitem[\protect\citeauthoryear{Thomas, Johansson  \& Maraston}{Thomas
  et~al.}{2011b}]{Thomas2011a}
Thomas D.,  Johansson J.,   Maraston C.,  2011b, MNRAS, 412, 2199

\bibitem[\protect\citeauthoryear{{Trager}, {Worthey}, {Faber}, {Burstein}  \&
  {Gonz{\'a}lez}}{{Trager} et~al.}{1998}]{Trager1998}
{Trager} S.~C.,  {Worthey} G.,  {Faber} S.~M.,  {Burstein} D.,   {Gonz{\'a}lez}
  J.~J.,  1998, \mn@doi [\apjs] {10.1086/313099}, \href
  {http://adsabs.harvard.edu/abs/1998ApJS..116....1T} {116, 1}

\bibitem[\protect\citeauthoryear{{Trager}, {Faber}, {Worthey}  \&
  {Gonz{\'a}lez}}{{Trager} et~al.}{2000}]{Trager2000}
{Trager} S.~C.,  {Faber} S.~M.,  {Worthey} G.,   {Gonz{\'a}lez} J.~J.,  2000,
  \mn@doi [\aj] {10.1086/301299}, \href
  {http://adsabs.harvard.edu/abs/2000AJ....119.1645T} {119, 1645}

\bibitem[\protect\citeauthoryear{{Vaughan}, {Davies}, {Zieleniewski}  \&
  {Houghton}}{{Vaughan} et~al.}{2018}]{Vaughan2017}
{Vaughan} S.~P.,  {Davies} R.~L.,  {Zieleniewski} S.,   {Houghton} R.~C.~W.,
  2018, \mn@doi [\mnras] {10.1093/mnras/stx3199}, \href
  {http://adsabs.harvard.edu/abs/2018MNRAS.475.1073V} {475, 1073}

\bibitem[\protect\citeauthoryear{{Vincenzo}, {Belfiore}, {Maiolino},
  {Matteucci}  \& {Ventura}}{{Vincenzo} et~al.}{2016}]{Vincenzo2016}
{Vincenzo} F.,  {Belfiore} F.,  {Maiolino} R.,  {Matteucci} F.,   {Ventura} P.,
   2016, \mn@doi [\mnras] {10.1093/mnras/stw532}, \href
  {https://ui.adsabs.harvard.edu/#abs/2016MNRAS.458.3466V} {458, 3466}

\bibitem[\protect\citeauthoryear{{Wake} et~al.,}{{Wake}
  et~al.}{2017}]{Wake2017}
{Wake} D.~A.,  et~al., 2017, \mn@doi [\aj] {10.3847/1538-3881/aa7ecc}, \href
  {http://adsabs.harvard.edu/abs/2017AJ....154...86W} {154, 86}

\bibitem[\protect\citeauthoryear{Walt, Colbert  \& Varoquaux}{Walt
  et~al.}{2011}]{VanDerWalt2011}
Walt S. V.~D.,  Colbert S.~C.,   Varoquaux G.,  2011, preprint (\mn@eprint
  {arXiv} {arXiv:1102.1523})

\bibitem[\protect\citeauthoryear{{Willett} et~al.,}{{Willett}
  et~al.}{2013}]{Willett2013}
{Willett} K.~W.,  et~al., 2013, \mn@doi [\mnras] {10.1093/mnras/stt1458}, \href
  {http://adsabs.harvard.edu/abs/2013MNRAS.435.2835W} {435, 2835}

\bibitem[\protect\citeauthoryear{{Woosley} \& {Weaver}}{{Woosley} \&
  {Weaver}}{1995}]{Woosley1995}
{Woosley} S.~E.,  {Weaver} T.~A.,  1995, \mn@doi [\apjs] {10.1086/192237},
  \href {http://adsabs.harvard.edu/abs/1995ApJS..101..181W} {101, 181}

\bibitem[\protect\citeauthoryear{Worthey, Faber, Gonzalez  \& Burstein}{Worthey
  et~al.}{1994}]{Worthey1994}
Worthey G.,  Faber S.~M.,  Gonzalez J.~J.,   Burstein D.,  1994, A{\&}A Suppl.
  Ser., 94, 687

\bibitem[\protect\citeauthoryear{{Worthey}, {Tang}  \& {Serven}}{{Worthey}
  et~al.}{2014}]{Worthey2014}
{Worthey} G.,  {Tang} B.,   {Serven} J.,  2014, \mn@doi [\apj]
  {10.1088/0004-637X/783/1/20}, \href
  {https://ui.adsabs.harvard.edu/#abs/2014ApJ...783...20W} {783, 20}

\bibitem[\protect\citeauthoryear{{Yan} et~al.,}{{Yan} et~al.}{2016a}]{Yan2016a}
{Yan} R.,  et~al., 2016a, \mn@doi [\aj] {10.3847/0004-6256/151/1/8}, \href
  {http://adsabs.harvard.edu/abs/2016AJ....151....8Y} {151, 8}

\bibitem[\protect\citeauthoryear{{Yan} et~al.,}{{Yan} et~al.}{2016b}]{Yan2016b}
{Yan} R.,  et~al., 2016b, \mn@doi [\aj] {10.3847/0004-6256/152/6/197}, \href
  {http://adsabs.harvard.edu/abs/2016AJ....152..197Y} {152, 197}

\bibitem[\protect\citeauthoryear{{Zieleniewski}, {Houghton}, {Thatte}, {Davies}
   \& {Vaughan}}{{Zieleniewski} et~al.}{2017}]{Zieleniewski2017}
{Zieleniewski} S.,  {Houghton} R.~C.~W.,  {Thatte} N.,  {Davies} R.~L.,
  {Vaughan} S.~P.,  2017, \mn@doi [\mnras] {10.1093/mnras/stw2712}, \href
  {http://adsabs.harvard.edu/abs/2017MNRAS.465..192Z} {465, 192}

\bibitem[\protect\citeauthoryear{{van Dokkum}, {Conroy}, {Villaume}, {Brodie}
  \& {Romanowsky}}{{van Dokkum} et~al.}{2017}]{vanDokkum2016}
{van Dokkum} P.,  {Conroy} C.,  {Villaume} A.,  {Brodie} J.,   {Romanowsky}
  A.~J.,  2017, \mn@doi [\apj] {10.3847/1538-4357/aa7135}, \href
  {http://adsabs.harvard.edu/abs/2017ApJ...841...68V} {841, 68}

\makeatother
\end{thebibliography}


\appendix
\section{Index measurements and parameters}
\label{sec:app_tables}
We provide our index measurements for the optical indices CN1, Ca4227, Fe4531, and C$_2$4668 for all radial bins in the mass ranges $9.9 - 10.2\;\log M/M_{\odot}$, $10.2 - 10.5\;\log M/M_{\odot}$ \& $10.5 - 10.8\;\log M/M_{\odot}$ in Table~\ref{tab:ind_measurements}, corrected to MILES resolution. The remaining indices H$\beta$, Mg$b$, Fe5270, Fe5335, and NaD were provided in \citetalias{Parikh2018}.

Also provided are the following parameters at each radial bin: age, metallicity, and the individual element abundances for C, N, Na, Mg, Ca, and Ti, derived using the TMJ models in Table~\ref{tab:parameters}.
\begin{table*}
  \centering
  \caption{Measured equivalent widths at all radial bins from $0 - 1\;R_\mathrm{e}$ for the mass bins. Indices have been corrected to MILES resolution. H$\beta$, Mg$b$, Fe5270, Fe5335, and NaD were provided in \citetalias{Parikh2018}}
  \label{tab:ind_measurements}
  \resizebox{\linewidth}{!}{%
  \begin{tabular}{lcccccccccc}
  \hline
  $10^{9.9 - 10.2}\;M_{\odot}$ & 0.0-0.1 & 0.1-0.2 & 0.2-0.3 & 0.3-0.4 & 0.4-0.5 & 0.5-0.6 & 0.6-0.7 & 0.7-0.8 & 0.8-0.9 & 0.9-1.0 \\
  \hline
  CN1 & 0.047$\pm$0.002 & 0.049$\pm$0.002 & 0.049$\pm$0.002 & 0.049$\pm$0.002 & 0.048$\pm$0.002 & 0.038$\pm$0.003 & 0.036$\pm$0.004 & 0.028$\pm$0.005 & 0.022$\pm$0.006 & 0.019$\pm$0.007 \\
  Ca4227 & 1.45$\pm$0.03 & 1.48$\pm$0.03 & 1.45$\pm$0.04 & 1.46$\pm$0.04 & 1.42$\pm$0.05 & 1.4$\pm$0.06 & 1.36$\pm$0.08 & 1.35$\pm$0.08 & 1.36$\pm$0.11 & 1.34$\pm$0.13 \\
  Fe4531 & 3.53$\pm$0.05 & 3.48$\pm$0.04 & 3.46$\pm$0.05 & 3.42$\pm$0.05 & 3.48$\pm$0.06 & 3.4$\pm$0.1 & 3.3$\pm$0.11 & 3.32$\pm$0.12 & 3.27$\pm$0.15 & 3.19$\pm$0.16 \\
  C$_2$4668 & 5.68$\pm$0.07 & 5.9$\pm$0.07 & 5.82$\pm$0.07 & 5.5$\pm$0.08 & 5.63$\pm$0.09 & 5.38$\pm$0.11 & 5.17$\pm$0.15 & 4.93$\pm$0.19 & 4.66$\pm$0.2 & 4.54$\pm$0.26 \\
  \hline
  $10^{10.2 - 10.5}\;M_{\odot}$ & 0.0-0.1 & 0.1-0.2 & 0.2-0.3 & 0.3-0.4 & 0.4-0.5 & 0.5-0.6 & 0.6-0.7 & 0.7-0.8 & 0.8-0.9 & 0.9-1.0 \\
  \hline
  CN1 & 0.083$\pm$0.001 & 0.082$\pm$0.002 & 0.071$\pm$0.002 & 0.064$\pm$0.002 & 0.059$\pm$0.002 & 0.054$\pm$0.003 & 0.048$\pm$0.003 & 0.044$\pm$0.005 & 0.04$\pm$0.005 & 0.033$\pm$0.009 \\
  Ca4227 & 1.51$\pm$0.03 & 1.5$\pm$0.03 & 1.47$\pm$0.03 & 1.46$\pm$0.03 & 1.46$\pm$0.04 & 1.45$\pm$0.05 & 1.43$\pm$0.07 & 1.43$\pm$0.08 & 1.36$\pm$0.1 & 1.32$\pm$0.15 \\
  Fe4531 & 3.59$\pm$0.04 & 3.6$\pm$0.04 & 3.64$\pm$0.05 & 3.66$\pm$0.05 & 3.61$\pm$0.06 & 3.54$\pm$0.08 & 3.53$\pm$0.1 & 3.48$\pm$0.13 & 3.45$\pm$0.13 & 3.32$\pm$0.18 \\
  C$_2$4668 & 6.33$\pm$0.05 & 6.17$\pm$0.05 & 5.99$\pm$0.05 & 5.85$\pm$0.08 & 5.68$\pm$0.09 & 5.72$\pm$0.12 & 5.47$\pm$0.14 & 5.46$\pm$0.18 & 5.4$\pm$0.23 & 5.22$\pm$0.27 \\
  \hline
  $10^{10.5 - 10.8}\;M_{\odot}$ & 0.0-0.1 & 0.1-0.2 & 0.2-0.3 & 0.3-0.4 & 0.4-0.5 & 0.5-0.6 & 0.6-0.7 & 0.7-0.8 & 0.8-0.9 & 0.9-1.0 \\
  \hline
  CN1 & 0.091$\pm$0.001 & 0.087$\pm$0.001 & 0.081$\pm$0.001 & 0.071$\pm$0.001 & 0.064$\pm$0.002 & 0.06$\pm$0.003 & 0.055$\pm$0.004 & 0.05$\pm$0.005 & 0.045$\pm$0.006 & 0.041$\pm$0.008 \\
  Ca4227 & 1.57$\pm$0.01 & 1.55$\pm$0.02 & 1.51$\pm$0.02 & 1.52$\pm$0.02 & 1.5$\pm$0.04 & 1.46$\pm$0.05 & 1.45$\pm$0.07 & 1.44$\pm$0.1 & 1.38$\pm$0.12 & 1.39$\pm$0.16 \\
  Fe4531 & 3.84$\pm$0.02 & 3.69$\pm$0.03 & 3.69$\pm$0.03 & 3.67$\pm$0.04 & 3.62$\pm$0.06 & 3.54$\pm$0.08 & 3.5$\pm$0.11 & 3.53$\pm$0.13 & 3.43$\pm$0.16 & 3.44$\pm$0.21 \\
  C$_2$4668 & 6.62$\pm$0.04 & 6.79$\pm$0.04 & 6.59$\pm$0.04 & 6.16$\pm$0.07 & 6.1$\pm$0.1 & 5.93$\pm$0.11 & 5.82$\pm$0.14 & 5.59$\pm$0.2 & 5.43$\pm$0.23 & 5.26$\pm$0.31 \\
  \hline
  \end{tabular}}
\end{table*}

\begin{table*}
  \centering
  \caption{Stellar population parameters and errors derived using the TMJ models and a combination of optical indices.}
  \label{tab:parameters}
  \resizebox{\linewidth}{!}{%
  \begin{tabular}{lllllllllll}
  \hline
    log Age (Gyr)        & 0-0.1                 & 0.1-0.2               & 0.2-0.3               & 0.3-0.4              & 0.4-0.5              & 0.5-0.6               & 0.6-0.7              & 0.7-0.8              & 0.8-0.9               & 0.9-1.0              \\
    \hline
    $9.9 - 10.2$  & 0.87$\pm$0.03 & 0.91$\pm$0.02 & 0.86$\pm$0.04 & 0.86$\pm$0.05 & 0.83$\pm$0.04 & 0.81$\pm$0.08 & 0.79$\pm$0.1 & 0.75$\pm$0.06 & 0.75$\pm$0.09 & 0.73$\pm$0.1 \\
    $10.2 - 10.5$ & 1.02$\pm$0.01 & 0.99$\pm$0.02 & 0.97$\pm$0.02 & 0.92$\pm$0.02 & 0.96$\pm$0.04 & 0.99$\pm$0.04 & 0.96$\pm$0.07 & 0.94$\pm$0.06 & 0.85$\pm$0.1 & 0.91$\pm$0.09 \\
    $10.5 - 10.8$ & 1.03$\pm$0.01 & 1.05$\pm$0.01 & 1.02$\pm$0.01 & 0.99$\pm$0.03 & 0.98$\pm$0.04 & 0.97$\pm$0.05 & 0.98$\pm$0.06 & 0.97$\pm$0.09 & 0.95$\pm$0.08 & 0.95$\pm$0.1 \\
      \hline
    [Z/H] (solar scaled)        & 0-0.1                 & 0.1-0.2               & 0.2-0.3               & 0.3-0.4              & 0.4-0.5              & 0.5-0.6               & 0.6-0.7              & 0.7-0.8              & 0.8-0.9               & 0.9-1.0              \\
      \hline
	$9.9-10.2$ & 0.033$\pm$0.016 & -0.0$\pm$0.018 & 0.015$\pm$0.019 & 0.004$\pm$0.022 & 0.001$\pm$0.023 & -0.015$\pm$0.034 & -0.031$\pm$0.043 & -0.034$\pm$0.04 & -0.035$\pm$0.056 & -0.048$\pm$0.064 \\
    $10.2 - 10.5$ & 0.072$\pm$0.014 & 0.087$\pm$0.018 & 0.085$\pm$0.016 & 0.093$\pm$0.017 & 0.034$\pm$0.024 & 0.007$\pm$0.03 & 0.004$\pm$0.043 & -0.008$\pm$0.046 & 0.02$\pm$0.046 & -0.027$\pm$0.064 \\
    $10.5 - 10.8$ & 0.092$\pm$0.012 & 0.072$\pm$0.011 & 0.073$\pm$0.012 & 0.081$\pm$0.018 & 0.053$\pm$0.024 & 0.042$\pm$0.031 & 0.007$\pm$0.044 & -0.002$\pm$0.054 & -0.018$\pm$0.061 & -0.036$\pm$0.081 \\
	\hline
    [C/Fe]        & 0-0.1                 & 0.1-0.2               & 0.2-0.3               & 0.3-0.4              & 0.4-0.5              & 0.5-0.6               & 0.6-0.7              & 0.7-0.8              & 0.8-0.9               & 0.9-1.0              \\
    \hline
    $9.9 - 10.2$ & 0.21$\pm$0.01 & 0.26$\pm$0.01 & 0.26$\pm$0.02 & 0.24$\pm$0.02 & 0.24$\pm$0.02 & 0.25$\pm$0.02 & 0.23$\pm$0.03 & 0.22$\pm$0.03 & 0.19$\pm$0.03 & 0.2$\pm$0.04 \\
    $10.2 - 10.5$ & 0.25$\pm$0.01 & 0.25$\pm$0.01 & 0.22$\pm$0.01 & 0.23$\pm$0.01 & 0.24$\pm$0.01 & 0.23$\pm$0.02 & 0.19$\pm$0.02 & 0.2$\pm$0.03 & 0.22$\pm$0.04 & 0.22$\pm$0.04 \\
    $10.5 - 10.8$ & 0.29$\pm$0.01 & 0.31$\pm$0.01 & 0.3$\pm$0.01 & 0.27$\pm$0.01 & 0.27$\pm$0.02 & 0.27$\pm$0.02 & 0.28$\pm$0.03 & 0.28$\pm$0.04 & 0.29$\pm$0.05 & 0.28$\pm$0.06 \\
	\hline
	[N/Fe]         & 0-0.1                 & 0.1-0.2               & 0.2-0.3               & 0.3-0.4              & 0.4-0.5              & 0.5-0.6               & 0.6-0.7              & 0.7-0.8              & 0.8-0.9               & 0.9-1.0              \\
    \hline
    $9.9 - 10.2$ & 0.0$\pm$0.02 & -0.01$\pm$0.01 & 0.02$\pm$0.02 & 0.11$\pm$0.02 & 0.06$\pm$0.02 & 0.03$\pm$0.02 & 0.08$\pm$0.03 & 0.07$\pm$0.03 & 0.06$\pm$0.05 & 0.09$\pm$0.05 \\
    $10.2 - 10.5$ & 0.18$\pm$0.01 & 0.24$\pm$0.01 & 0.17$\pm$0.01 & 0.16$\pm$0.01 & 0.13$\pm$0.02 & 0.03$\pm$0.02 & 0.03$\pm$0.03 & 0.0$\pm$0.04 & 0.01$\pm$0.05 & -0.03$\pm$0.05 \\
    $10.5 - 10.8$ & 0.19$\pm$0.01 & 0.13$\pm$0.01 & 0.14$\pm$0.01 & 0.15$\pm$0.01 & 0.08$\pm$0.02 & 0.09$\pm$0.03 & 0.06$\pm$0.04 & 0.05$\pm$0.04 & 0.08$\pm$0.05 & 0.06$\pm$0.07 \\
	\hline
    [Na/Fe]        & 0-0.1                 & 0.1-0.2               & 0.2-0.3               & 0.3-0.4              & 0.4-0.5              & 0.5-0.6               & 0.6-0.7              & 0.7-0.8              & 0.8-0.9               & 0.9-1.0              \\
    \hline
    $9.9 - 10.2$ & 0.25$\pm$0.01 & 0.28$\pm$0.01 & 0.27$\pm$0.02 & 0.26$\pm$0.02 & 0.23$\pm$0.02 & 0.22$\pm$0.02 & 0.19$\pm$0.02 & 0.17$\pm$0.03 & 0.12$\pm$0.03 & 0.12$\pm$0.04 \\
    $10.2 - 10.5$ & 0.37$\pm$0.01 & 0.38$\pm$0.01 & 0.32$\pm$0.01 & 0.32$\pm$0.01 & 0.28$\pm$0.01 & 0.23$\pm$0.02 & 0.2$\pm$0.02 & 0.18$\pm$0.03 & 0.18$\pm$0.04 & 0.17$\pm$0.04 \\
    $10.5 - 10.8$ & 0.47$\pm$0.01 & 0.44$\pm$0.01 & 0.41$\pm$0.01 & 0.37$\pm$0.01 & 0.34$\pm$0.02 & 0.29$\pm$0.02 & 0.28$\pm$0.03 & 0.28$\pm$0.04 & 0.28$\pm$0.05 & 0.26$\pm$0.06 \\
    \hline
    [Mg/Fe]        & 0-0.1                 & 0.1-0.2               & 0.2-0.3               & 0.3-0.4              & 0.4-0.5              & 0.5-0.6               & 0.6-0.7              & 0.7-0.8              & 0.8-0.9               & 0.9-1.0              \\
    \hline
    $9.9 - 10.2$ & 0.22$\pm$0.01 & 0.23$\pm$0.01 & 0.25$\pm$0.02 & 0.26$\pm$0.02 & 0.23$\pm$0.02 & 0.26$\pm$0.02 & 0.25$\pm$0.02 & 0.26$\pm$0.02 & 0.26$\pm$0.03 & 0.27$\pm$0.04 \\
    $10.2 - 10.5$ & 0.26$\pm$0.01 & 0.28$\pm$0.01 & 0.26$\pm$0.01 & 0.27$\pm$0.01 & 0.28$\pm$0.01 & 0.24$\pm$0.02 & 0.22$\pm$0.02 & 0.21$\pm$0.02 & 0.25$\pm$0.04 & 0.25$\pm$0.04 \\
    $10.5 - 10.8$ & 0.29$\pm$0.01 & 0.28$\pm$0.0 & 0.28$\pm$0.01 & 0.3$\pm$0.01 & 0.28$\pm$0.01 & 0.28$\pm$0.02 & 0.29$\pm$0.03 & 0.3$\pm$0.03 & 0.32$\pm$0.04 & 0.31$\pm$0.06 \\
    \hline
    [Ca/Fe]        & 0-0.1                 & 0.1-0.2               & 0.2-0.3               & 0.3-0.4              & 0.4-0.5              & 0.5-0.6               & 0.6-0.7              & 0.7-0.8              & 0.8-0.9               & 0.9-1.0              \\
    \hline
    $9.9 - 10.2$ & 0.07$\pm$0.02 & 0.15$\pm$0.01 & 0.15$\pm$0.02 & 0.15$\pm$0.02 & 0.16$\pm$0.02 & 0.14$\pm$0.02 & 0.12$\pm$0.03 & 0.11$\pm$0.03 & 0.05$\pm$0.04 & 0.06$\pm$0.05 \\
    $10.2 - 10.5$ & 0.08$\pm$0.01 & 0.08$\pm$0.01 & 0.02$\pm$0.01 & 0.03$\pm$0.01 & 0.07$\pm$0.02 & 0.06$\pm$0.02 & 0.02$\pm$0.03 & 0.06$\pm$0.03 & 0.03$\pm$0.05 & 0.02$\pm$0.05 \\
    $10.5 - 10.8$ & 0.12$\pm$0.01 & 0.1$\pm$0.01 & 0.1$\pm$0.01 & 0.08$\pm$0.01 & 0.1$\pm$0.02 & 0.08$\pm$0.02 & 0.12$\pm$0.03 & 0.11$\pm$0.04 & 0.1$\pm$0.05 & 0.09$\pm$0.06 \\
    \hline
    [Ti/Fe]        & 0-0.1                 & 0.1-0.2               & 0.2-0.3               & 0.3-0.4              & 0.4-0.5              & 0.5-0.6               & 0.6-0.7              & 0.7-0.8              & 0.8-0.9               & 0.9-1.0              \\
    \hline
    $9.9 - 10.2$ & 0.17$\pm$0.02 & 0.16$\pm$0.02 & 0.19$\pm$0.02 & 0.19$\pm$0.03 & 0.19$\pm$0.03 & 0.2$\pm$0.04 & 0.15$\pm$0.05 & 0.19$\pm$0.05 & 0.16$\pm$0.06 & 0.14$\pm$0.07 \\
    $10.2 - 10.5$ & 0.18$\pm$0.01 & 0.22$\pm$0.02 & 0.22$\pm$0.02 & 0.27$\pm$0.02 & 0.26$\pm$0.03 & 0.18$\pm$0.03 & 0.16$\pm$0.04 & 0.13$\pm$0.05 & 0.18$\pm$0.07 & 0.12$\pm$0.07 \\
    $10.5 - 10.8$ & 0.34$\pm$0.02 & 0.24$\pm$0.01 & 0.27$\pm$0.01 & 0.28$\pm$0.02 & 0.26$\pm$0.03 & 0.23$\pm$0.04 & 0.23$\pm$0.05 & 0.27$\pm$0.07 & 0.27$\pm$0.09 & 0.27$\pm$0.12 \\
    \hline
    \end{tabular}}
\end{table*}

\bsp	
\label{lastpage}
\end{document}